\documentclass[10pt]{article}
\usepackage[utf8]{inputenc}
\usepackage{times}
\usepackage{graphicx,latexsym,amssymb}
\makeatletter
  \renewcommand\@biblabel[1]{}
\makeatother

  \newcommand{\bq}{\begin{quote}} \newcommand{\eq}{\end{quote}}     \newcommand{\M}{\mid\mid} \newtheorem{Th}{Theorem}   \newtheorem{ax}{Axiom}   \newtheorem{lm}{Lemma}  \newtheorem{df}{Definition}     \newtheorem{pr}{Proposition}  \newtheorem{cl}{Corollary}   \newtheorem{re}{Remark}     \newtheorem{as}{Assumption}   \newtheorem{wg}{Wild Guess} \newtheorem{ex}{Example} \newcommand{\bth}{\begin{Th}\hspace{-5pt}{\bf .} \ }  \newcommand{\Eth}{\end{Th}} \newcommand{\bax}{\begin{ax}\hspace{-5pt}{\bf .} \ }  \newcommand{\eax}{\end{ax}} \newcommand{\blm}{\begin{lm}\hspace{-5pt}{\bf .} \ } \newcommand{\elm}{\end{lm}} \newcommand{\bdf}{\begin{df}\hspace{-5pt}{\bf .} \ }    \newcommand{\edf}{\end{df}}  \newcommand{\bpr}{\begin{pr}\hspace{-5pt}{\bf .} \ }  \newcommand{\epr}{\end{pr}} \newcommand{\bcl}{\begin{cl}\hspace{-5pt}{\bf .} \ }  \newcommand{\ecl}{\end{cl}} \newcommand{\bre}{\begin{re}\hspace{-5pt}{\bf .} \ } \newcommand{\ere}{\end{re}} \newcommand{\bas}{\begin{as}\hspace{-5pt}{\bf .} \ } \newcommand{\eas}{\end{as}} \newcommand{\bwg}{\begin{wg}\hspace{-5pt}{\bf .} \ } \newcommand{\ewg}{\end{wg}} \newcommand{\bex}{\begin{ex}\hspace{-5pt}{\bf .} \ }   \newcommand{\eex}{\end{ex}}   \newcommand{\bit}{\begin{itemize}} \newcommand{\eit}{\end{itemize}\par\noindent} \newcommand{\beq}{\begin{equation}}  \newcommand{\eeq}{\end{equation}\par\noindent} \newcommand{\beqa}{\begin{eqnarray*}} \newcommand{\eeqa}{\end{eqnarray*}\par\noindent} \newcommand{\beqn}{\begin{eqnarray}}   \newcommand{\eeqn}{\end{eqnarray}\par\noindent}
\newcommand{\diraco}{\mid 0 \rangle}
\newcommand{\dirac}{\mid 1 \rangle}
\newcommand{\Odiraco}{\overline{\mid 0 \rangle}}
\newcommand{\Odirac}{\overline{\mid 1 \rangle}}

\title{LQP: The Dynamic Logic of Quantum Information}
\author{Alexandru Baltag\footnote{Oxford University Computing Laboratory, UK}
 \, and Sonja Smets\footnote{Vrije Universiteit Brussel, Flanders' Fund for Scientific Research Post-doc, Belgium}}
\date{}
\begin{document}
\maketitle


\begin{abstract}
\par\noindent
The main contribution of this paper is the introduction of a {\it dynamic logic} formalism for reasoning about information flow in {\it composite} quantum systems. This builds on our
previous work on a {\it complete} quantum dynamic logic for {\it single} systems. Here we extend
that work to a {\it sound} (but not necessarily complete) logic for composite systems, which
brings together ideas from the quantum logic tradition with concepts from (dynamic) modal
logic and from quantum computation. This {\it Logic of Quantum Programs (LQP)} is capable of
expressing important features of quantum measurements and unitary evolutions of
multi-partite states, as well as giving logical characterisations to various forms of
entanglement (for example, the Bell states, the {\it GHZ} states etc.). We present a finitary syntax, a relational semantics and a sound proof system for this logic. As applications, we use our
system to give formal correctness proofs for the Teleportation protocol and for a standard
Quantum Secret Sharing protocol; a whole range of other quantum circuits and programs,
including other well-known protocols (for example, superdense coding, entanglement
swapping, logic-gate teleportation etc.), can be similarly verified using our logic.

\end{abstract}

\section{Introduction}
As a natural extension of Hoare Logic, Propositional Dynamic Logic ($PDL$) is an important tool for the
logical study of programs, especially by providing a basis for {\it program verification}. In the context
of recent advances in quantum programming, it is natural to look for a
{\it quantum} version of $PDL$, which could play the same role
in proving correctness for quantum programs that classical $PDL$ (and Hoare logic) played for classical
programs.

The search for such a `quantum $PDL$' has been one of the main objectives of our previous investigation into the logic of quantum information flow. In a series of presentations \cite{Baltag,Smets} and papers \cite{BaltagSmets, LICS,Portugal}, we have proposed several logical systems: in \cite{BaltagSmets} we focused on {\it single systems}\footnote{A {\it single} system is just an isolated physical system; the possible states of such a system are represented in quantum mechanics as rays in some Hilbert space. By contrast, a {\it composite} (also called {\it compound}, or {\it multi-partite}) system is one that we can think of as being composed of two (or more) distinct physical (sub)systems. The corresponding Hilbert space is the {\it tensor product} of each of the spaces associated to the subsystems. So, in a sense, single systems {\it subsume} composite systems (since any tensor product of Hilbert
spaces is just another Hilbert space). However, treating a system as being composite amounts to having a more detailed complex theory of the system (compared with treating it as a single system) - a theory that captures the specific features arising from being a {\it composite} structure, {\it in addition} to the general features of any physical system. A `logic' for compound systems will thus be a {\it richter} logic than one for single systems.}
 and presented two equivalent {\it complete axiomatizations} for a
Logic of Quantum Actions, $LQA$, which allows actions such as measurements and unitary evolutions, but no
entanglements.  The completeness result was obtained with respect to
infinite-dimensional classical Hilbert spaces, as models for {\it single quantum systems}. The challenge of providing
a similar axiomatization for {\it compound systems} was taken up in
\cite{LICS}, where a first proposal for a {\it logic of multi-partite quantum systems} was sketched.

\medskip

In this paper we elaborate further, simplify and improve on the work outlined in \cite{LICS}, and
develop a full-fledged {\it Logic of Quantum Programs} $LQP$.\footnote{But note the difference between our
logic $LQP$ and the approach with a similar name in \cite{Brunet}:
our dynamic logic goes much further in capturing essential properties of quantum systems and quantum programs,
as well as in recovering the ideas of traditional quantum logic (see e.g. \cite{DC,DC2,G}).}
This includes:
\begin{enumerate}
\item A simple {\it finitary syntax} for a
{\it modal language}, based on a minor variation of classical $PDL$, with dynamic modalities corresponding
to (weakest preconditions of) quantum programs.
\item A
{\it relational semantics} for this logic, in terms
of {\it quantum states and quantum actions over a finite-dimensional Hilbert space}.
\item A {\it sound (but not necessarily complete\footnote{Unlike the case of infinite-dimensional single systems, for which a complete logic was given in Baltag and Smets [2005a], the problem of finding a complete proof system for the logic $LQP$ is still open.}) proof system}, which includes axioms to handle {\it separation}, {\it locality} and {\it entanglement}.
\item {\it Formal proofs (in our proof system $LQP$)
of non-trivial computational properties of compound quantum systems}.
\item
 An {\it analysis} (with a {\it formal correctness proof}) of the {\it Teleportation} and {\it quantum secret sharing protocols}.
\end{enumerate}

\medskip\par\noindent
More generally, the strength of $LQP$ lies in the fact that it can provide {\it fully formal correctness proofs for a whole class of quantum circuits and protocols}, a class that includes logic-gate teleportation, superdense coding and entanglement swapping, as well as more complex circuits built using quantum gates and measurements.

The logic introduced here brings together a number of ideas from several fields:
theoretical foundations of quantum mechanics, operational quantum logic, dynamic modal
logic, spatial logic and quantum computation. In the rest of this section we give an overview
of the main concepts underlying the logic $LQP$.

The first fundamental idea of our approach connects two independent lines of research. The first is the long tradition in the logical-algebraic foundations of quantum mechanics, which, in particular, has produced various `dynamic' interpretation of quantum logic (QL) in \cite{Daniel1,Daniel2,Faure,Amira,Compoundness,Quantale,LogicOfDynamics}, \cite{SasakiIsNot,CS,OnCausation,
MalPaper}. The second line is the work on modal `action' logics in Computer Science, the main example being Dynamic Logic ($PDL$) and its relatives (Hoare logic, but also dynamic interpretations of basic modal logics as languages for
`processes' or labelled transition systems, for example, Hennessey-Milner logic).

We stress the fact that, until our recent work \cite{BaltagSmets}, these two
traditions were not only independent, but did not even share a common language. The
use of the word `dynamic' in the $QL$ tradition did not have much in common with
`dynamic' logic; $QL$ aimed for an algebraic axiomatisation of quantum systems based on
the {\it non-distributive lattice of `quantum properties'}, structure obtained by abstracting away
from the lattice of {\it projectors in a Hilbert space ${\cal H}$} (or, equivalently, the lattice of closed
linear subspaces of ${\cal H}$); the goal was to obtain representation theorems for these logical
structures with respect to Hilbert spaces, thus allowing one to claim a `rational', `logical'
(or `operational') reconstruction of quantum mechanics\footnote{This goal was partially realised in \cite{JP3} and \cite{JP4}, and later improved on in \cite{Soler} and \cite{Mayet}, and related work.}. In this context, the `dynamic' twist has to do with the addition of features belonging to {\it physical dynamics} to the standard (static) $QL$ description:
\begin{itemize}
\item First, a `dynamic' interpretation was given to the main structure (the lattice of
properties) and the logical connectives (quantum implication and quantum disjunction)
of $QL$: in, for example, \cite{OnCausation}, \\ \cite{SasakiIsNot} and \cite{LogicOfDynamics}
(and partially anticipated in \cite{Hardegree1,Hardegree2} and \\ \cite{Beltrametti})
the quantum-logical connectives are interpreted dynamically, as expressing
{\it potential causality} (that is, what in computer science is known as {\it weakest preconditions}).
\item Second, some of the researchers in $QL$ went on to incorporate `true' physical dynamics,
that is, Schrodinger flows (unitary evolutions), into the algebra as operators on the
underlying lattice; the resulting structure is a {\it quantale} of `quantum actions', which was
introduced and investigated in \\ \cite{Quantale} and \cite{CS}.
\end{itemize}

In contrast, the modal logician's (and the computer scientist's) use of `dynamics' refers to
modelling processes as {\it labelled transition systems} (Kripke models), in which the possible `actions' are represented as {\it binary relations} between possible states, and the natural
descriptive language is {\it modal}, having dynamic modalities to express {\it weakest preconditions} (ensuring given post-conditions after specific actions)\footnote{See, for example, \cite{Harel} for an introduction to dynamic modalities $[\pi]\psi$ describing weakest preconditions ensuring (the satisfaction of some post-condition) $\psi$ after the execution of action $\pi$.}. Thus, the `first fundamental idea' of
our logic, an idea first presented in \cite{Baltag} and \cite{Smets} and published in \cite{BaltagSmets}, is to connect these two traditions by giving a {\it quantum (re)interpretation
of Dynamic Logic}, in which {\it both (projective) measurements and unitary evolutions are
treated as modal actions}, and to use this formalism in order to improve on the known
representation theorems in $QL$. In this quantum reinterpretation, the
`test' actions $\varphi?$ of $PDL$ (which are used to capture conditional programs in dynamic logic) are to be read as {\it `successful measurements'
of a quantum property} $\varphi$ (that is, as projectors in a Hilbert space over the subspace generated by the set of states
satisfying $\varphi$), while the other basic actions of $PDL$ are taken to be {\it quantum gates}
(i.e. unitary operators on a Hilbert space). As shown in \cite{Baltag, Smets, BaltagSmets}, this immediately
allows us to re-capture in our (Boolean) logic all the power of traditional (non-Boolean) Quantum Logic: the
`quantum disjunction' (expressing superpositions), the `quantum negation' (the so-called `orthocomplement' $\sim\varphi$, which expresses the necessary failure of
a measurement) and the `quantum implication' (the so-called `Sasaki hook' $\phi \stackrel{S}{\to} \psi$, which captures causality in quantum measurements)
are all expressible using quantum-dynamic modalities $[\varphi?] \psi$
(which capture {\it weakest preconditions
 of quantum measurements}).\footnote{Indeed, it turns out that a quantum implication  $\phi \stackrel{S}{\to} \psi$
is simply equivalent to the weakest precondition $[\varphi?] \psi$.
In quantum logic, this dynamic view can be traced back
to the analysis of the Sasaki hook as a Stalnaker conditional presented in \cite{Hardegree1,Hardegree2}
and is reflected on in, for example, \cite{Beltrametti} and \cite{OnCausation}.}
In other words: in our logic (unlike other logical approaches to quantum systems),
{\it all the non-classical `quantum' effects are captured using a non-classical `logical dynamics',
while keeping the classical, Boolean structure of the underlying propositional logic of `static' properties}.

The second fundamental idea of our approach was originally outlined in \\ \cite{LICS}, and consists of adding {\it spatial features} to dynamic
logic, in order to capture relevant properties of {\it multi-partite (that is, compound) quantum systems}
(for example, separation, locality, entanglement). For this, we
use a finite set $N$ of {\it indices} to denote the most basic `parts' (qubits) of the system, and
use {\it sets of indices} $I\subseteq N$ to denote all the (possibly compound) subsystems; we have
special propositional constants $1$, $0$, $+$, and so on, to express the fact that qubits are in the state
$\mid 1\rangle$, $\mid 0\rangle$ or $\mid + \rangle$, and so on; we use a basic propositional formula
$\top_I$ to express {\it `separation'} \footnote{This can be compared with the
{\it exogenous quantum logic approach} in \cite{EQL}, which makes use of general modal operators to separate subsystems.}
(the fact that qubits in the subsystem $I$ are separated from the rest);
and we have a basic program $\top_I$, denoting a {\it non-determined (that is, randomly chosen) local transformation}
(affecting only the qubits in the subsystem $I$). These ingredients are enough to define all the relevant
spatial features we need, and in particular to define the notion of {\it (local) component $\varphi_I$ of a
(global) property $\varphi$}, the notion of {\it ($I$-)local property} $I(\varphi)$ (that is, $\varphi$ is a property
of the separated $I$-subsystem) and the notion of
{\it ($I$-)local program} $I(\pi)$ (that is, $\pi$ is a program affecting only the $I$-subsystem).

The third fundamental idea that underlies our approach comes from \\ \cite{Compoundness} and \cite{Coecke2}, and was further elaborated in a category-theore\-ti\-cal setting in \cite{AbramskiCoecke}): this is a {\it computational understanding of entanglement, in which an entangled state is seen as
a `static' encoding of a program}. Mathematically, this comes from the simple observation
that a tensor product $H_i \otimes H_j$ of two Hilbert spaces is canonically isomorphic to
the space $H_i \to H_j$ of all linear maps between the two spaces. But, as noted in \cite{Compoundness,Coecke2},
this isomorphism has a {\it physical meaning}: the entangled state $\overline{\pi_{ij}}$, which `encodes' (via the above isomorphism) the linear map $\pi: H_i \to H_j$, has the property that {\it any successful
measurement of its $i$-th qubit} (resulting in some local output-state $q_i$) {\it induces a correlative
collapse of the $j$-th qubit, whose local output-state} (after the collapse) {\it
is computed by the map $\pi$} (that is, it is given
by $\pi(q)_j$). So {\it the above isomorphism captures the correlations between possible results of potential
local measurements} (on the two qubits). We use this idea to define formulas $\overline{\pi_{ij}}$
that characterize such specifically entangled states (by using weakest preconditions to express potential
behavior under possible measurements). The fundamental correlation given by the above isomorphism is
then stated as our {\it `Entanglement Axiom'}, which plays a central role in our system.

This {\it combination of quantum-dynamic and spatial logic} is what allows
us to give {\it a logical characterization of Bell states} and of {\it various quantum gates},
and {\it to prove from our axioms for
highly non-trivial properties of quantum
information flow} (such as the `Teleportation Property, the
`Agreement Property', the `Entanglement Preparation' and `Entanglement Composition' lemmas etc.).

It is well-known that $PDL$, and
its fragment the Hoare Logic, are among the main logical formalisms used in {\it program verification} of classical
programs, i.e. in checking that a given (classical) program is {\it correct}
(in the sense of meeting the required specifications). It is thus natural
to expect our quantum dynamic logic to play a significant role in the formal {\it verification of quantum programs}.
In this paper, we partially fulfil this expectation by giving a {\it fully axiomatic correctness proof for the
Teleportation protocol and for a Quantum Secrete Sharing protocol}; more details, and similar proofs for other quantum programs (Logic-Gate Teleportation, Super-Dense Coding, Entanglement Swapping, and so on) can be found in \cite{Akatov}. More generally, our logic can be used for the formal verification of a whole range of quantum programs\footnote{Indeed, one may claim that {\it any} quantum circuit in which {\it probabilities do not play an essential role} can, in principle, be verified using our logic (or some trivial extension obtained by adding constants for other relevant states and logic gates).}, including all the circuits covered by the `entanglement networks' approach in \cite{Coecke2}.

Finally, we mention here some of the {\it limitations of our approach}, which arise from
our {\it purely qualitative, logic-based} view of quantum information.
The quantitative aspects are thus
neglected:
in our presentation, we follow the {\it operational quantum logic} tradition, as in, for example, \cite{JP1} and \cite{JP2}, by abstracting away from complex numbers, `phases' and probabilities.
As customary in quantum logic, we identify the `states' of a physical system
with {\it rays} in a Hilbert space\footnote{A ray is a one-dimensional
linear subspace.}, rather than with unitary vectors, and consequently, our
programs will be {\it `phase-free'}. This is a
serious
limitation, as phase aspects are important in quantum computation;
there are ways to re-introduce (relative) phases in our approach, but this gives rise to a much more
complicated logic, and so we will leave this development for future work.
Similarly, although our dynamic logic {\it cannot express probabilities,
but only `possibilities'} (via the dynamic modalities, which capture the system's potential
behavior under possible actions), {\it there exist natural extensions of this setting to a probabilistic modal
logic}. One of our projects is to work out the full details of this setting, developing a proof system for probabilistic
$LQP$.

\section{Preliminaries: quantum frames}
In this section we organize Hilbert spaces as relational structures, called {\it quantum frames}
(also called {\it quantum transition systems} in \cite{BaltagSmets}). We first
study the quantum frames of {\it single} quantum systems, then we consider systems {\it compound system}, that is, the quantum frames corresponding to the tensor products of Hilbert spaces, which represent physical systems that can be thought of as being composed of {\it parts} (subsystems). In this latter case we restrict our attention to {\it systems composed of finitely many `qubits'}.

\subsection{Single-system quantum frames}

A {\it modal frame} is a set of {\it states}, together with
a family of {\it binary relations} between states. A (generalised) {\it PDL frame}
is a modal frame $(\Sigma, \{\stackrel{S?}{\to}\}_{S \in {\cal L}},
\{ \stackrel{a}{\to} \}_{a \in {\cal A}})$,
in which the relations on the set of states $\Sigma$ are of two types:
the first, called {\it tests} and denoted by $S?$, are labelled with
subsets $S$ of $\Sigma$, coming from a given family ${\cal L}\subseteq
{\mathcal P}(\Sigma)$ of sets, called {\it testable properties}; the
others, called {\it actions}, are labelled with action labels $a$ from
a given set ${\cal A}$.

Given a $PDL$ frame, there exists a standard way
to give a semantics to the usual language of {\it Propositional
Dynamic Logic}. Classical $PDL$ can be considered as a special case
of such a logic, in which tests are given by {\it classical tests}:
$s \stackrel{S?}{\to} t$ if and only if $s = t\in S$.
Observe that {\it classical tests, if executable, do not change the
current state}.

In the context of quantum systems, a natural idea is to replace
classical tests by `quantum tests', given by {\it quantum
measurements}. Such tests will obviously change
the
state of the system. To model them, we introduce a special kind of $PDL$ frames: {\it quantum
frames}. The tests are essentially given by {\it projectors} in a
Hilbert space, while the other basic actions are given by {\it unitary evolutions}. In  \cite{BaltagSmets}, we considered $PDL$ with this non-standard
semantics, having essentially the same truth clauses as in the classical case, but
interpreted in quantum frames. What we obtained was a {\it `quantum PDL'}, in which the traditional (orthomodular) `quantum logic\footnote{See, for example, see e.g. \cite{DC,DC2} and \cite{G}.}' could be embedded as a fragment (corresponding to the {\it negation-free, test only}
part of quantum $PDL$). In this paper, we extend the syntax of this logic to deal with subsystems and entanglements.

Recall that
a {\it Hilbert space} ${\cal H}$ is a complex vector space with an inner product $\langle -\mid -\rangle$, which is complete in
the induced metric. The {\it adjoint} (or {\it Hermitian conjugate}) of a linear map
$F:{\cal H}\to {\cal H}$ is the unique linear map
$F^{\dagger}: {\cal H}\to {\cal H}$ s.t.
$\langle x\mid F(y)\rangle = \langle F^{\dagger}(x) \mid y \rangle$, for all $x, y\in {\cal H}$.
For any closed linear subspace $W\subseteq {\cal H}$, the {\it projector} $P_W: H\to H$ onto $W$ is given by:
$P_W(u+v)=u$, for all $u\in W, v\in W^{\perp}$. Projectors are linear, idempotent ($P\circ P=P$) and
self-adjoint ($P^{\dagger}=P$).
A {\it unitary transformation} is a linear map $U$ on ${\cal H}$ s.t. $U\circ U^{\dagger}=
U^{\dagger}\circ U=id$, where $id$ is the identity on ${\cal H}$. Unitary operators preserve inner products.

In Quantum Mechanics, projectors are used to represent {\it (successful) measurements}. A measurement is
in fact a {\it set} of projectors (over mutually orthogonal subspaces); but, whenever a measurement is successfully
performed, {\it only one} of the projectors is `actualised': the outcome is given by that particular projector.
In Quantum Mechanics, unitary transformations represent {\it reversible evolutions} of a system.
In Quantum Computation, they correspond
to {\it quantum-logical gates}.

\bigskip
\par\noindent\textbf{Quantum frames}
\par\noindent
Given a Hilbert space ${\cal H}$, the following steps construct a
{\it Quantum (PDL) Frame}

$$ {\Sigma({\cal H})} :=  {(\Sigma,
\{\stackrel{S?}{\to}\}_{S \in {\cal L}}, \{ \stackrel{U}{\to} \}_{U \in {\cal U}})}$$

\begin{enumerate}
\item Let $\Sigma$ be the set of {\it one dimensional subspaces} of
${\cal H}$, called the set of {\it states}. We denote a state
$s=\overline{x}$ of ${\cal H}$ using any of the non-zero vectors $x
\in {\cal H}$ that generate it, as a subspace. Note that any two vectors
that differ only in {\it phase} (that is, $x=\lambda y$, with $\lambda\in
C$ with $|\lambda|=1$) will generate the same state
$\overline{x}=\overline{y}\in \Sigma$.
\item We call two states $s$ and $t$ in $\Sigma$ {\it orthogonal}, and
write $s \perp t$, if every two vectors $x\in s, y\in t$ are orthogonal, that is, if
$\forall x \in s \forall  y \in t \, \langle x
\mid y \rangle = 0$. Equivalently, we can state that $s \perp t$ iff $\exists x  \in s,  y \in t$ with $x  \not = 0$, $y \not
= 0$ and $ \langle x \mid y \rangle = 0$.
We put $S^{\perp} := \{t \in \Sigma \mid t \perp s \mbox{ for \, all }s
\in S \}$; and we denote by $\overline{S}=S^{\perp \perp} := (S^{\perp})^{\perp}$ the
biorthogonal closure of $S$. In particular, for a singleton
$\{x\}$, we just write $\overline{x}$ for $\overline{\{x\}}$, which
agrees with the notation $\overline{x}$ used above to denote the state
generated by $x$.
\item A set of states $S \subseteq \Sigma$ is called a {\it (quantum) testable
property} iff it is
{\it biorthogonally
closed}, i.e. if $\overline{S} = S$. (Note that $S \subseteq \overline{S}$ is always the case.)
We use ${\cal L}\subseteq P(\Sigma)$ to denote the family of all quantum testable
properties. All the {\it other} sets $S\in P(\Sigma)\setminus {\cal
L}$ are called {\it non-testable properties}.
\item There is a natural bijective correspondence between the family
${\cal L}$ of all testable
properties and the family ${\cal W}$ of all {\it closed linear
subspaces} $W$ of ${\cal H}$, the bijection being given by
$S \, \, \mapsto \, \, W_S=:\bigcup S$. Observe that, under this
correspondence,
the image of the biorthogonal closure $\overline{S}$ of any arbitrary set
$S\subseteq \Sigma$ is the closed linear subspace $\overline{\bigcup
S}\subseteq {\cal H}$ generated by the union $\bigcup S$ of all states
in $S$.
\item For each testable property $S\in {\cal L}$, there exists a
partial map $S?$ on $\Sigma$, called a {\it quantum test}. If
$W= W_S=\bigcup S$ is the corresponding subspace of ${\cal H}$, then
the quantum test is the map induced on states by the {\it projector}
$P_W$ onto the subspace $W$. In other words, it is given by:
\begin{eqnarray*}
S?(\overline{x}) & := & \overline{P_W(x)} \in \Sigma\, , \,    \mbox{
if } \overline{x} \not \in S^\perp \,  ( \mbox{ i.e.  if } P_W(x) \not
= 0) \\
S?(\overline{x}) & := & \mbox{ undefined }\, , \, \mbox{ otherwise }.
\end{eqnarray*}
We use $\stackrel{S?}{\to} \subseteq \Sigma \times \Sigma$ to denote the
binary relation corresponding to the partial map $S?$ that is given by:
$s \stackrel{S?}{\to} t$ if and only if $S?(s) = t$. So we have
{\it a family of binary relations indexed by the testable properties}
$S \in {\cal L}$.

\item For each unitary transformation $U$ on ${\cal H}$, consider the
corresponding binary relation $\stackrel{U}{\to} \subseteq \Sigma
\times \Sigma$, given by:  $s \stackrel{U}{\to} t$ if and only if
$ U (x)  =  y $ for some  non-zero vectors $x \in s, y  \in t$. So we obtain
{\it a family of binary relations indexed by the unitary
transformations} $U \in {\cal U}$ (where ${\cal U}$ is the set of
unitary transformations on ${\cal H}$).
\end{enumerate}

So a quantum frame is just a $PDL$ frame built on top of a given Hilbert
space ${\cal H}$, by taking one-dimensional subspaces as `states', projectors as `tests' and unitary evolutions
as `actions'.
Our notion of `state' in this paper
is closely connected to the way quantum logicians approach quantum
systems. As mentioned in the Introduction, this
imposes some limits to our approach - mainly that we will not be able
to express {\it phase}-related properties.

\bigskip\par\noindent
{\bf Operators on states, adjoints and generalised tests}
\par\noindent
To generalise the notation we introduced earlier, observe that every
{\it linear operator} $F: {\cal H} \to {\cal H}$ induces
a partial map $F: {\Sigma} \to {\Sigma}$ on states (that is, subspaces), given by
$F(\overline{x}) = \overline{F(x)} $, if $F(x)\not=0$ (and undefined, otherwise). (Note that {\it linearity}
ensures that this map on states is well-defined.) In particular, every
map $F:\Sigma \to \Sigma$ obtained in this way has an {\it adjoint}
$F^{\dagger}:\Sigma \to \Sigma$, defined as the map on states induced by
the adjoint of the linear operator $F$ on
${\cal H}$. Observe that, for unitary transformations $U$, the adjoint
is the inverse: $U^{\dagger}=U^{-1}$
Also, one can
naturally generalise {\it quantum tests} to arbitrary, possibly {\it
non-testable properties}, $S\subseteq \Sigma$, by putting: $S ? :=
\overline{S} ?$. So we identify a test of a `non-testable' property $S$
with the quantum test of its biorthogonal closure. Observe that $S?^{\dagger}=S?$
(since projectors are self-adjoint).

\smallskip\par\noindent\textbf{Measurement (non-orthogonality) relation}
\par\noindent
For all $s,t \in \Sigma$, let $s \to t$ if and only if $s
\stackrel{S?}{\to} t$ for some property $S \in {\cal L}$.
In other words, $s\to t$ means that one can reach state $t$ by doing {\it
some measurement} on state $s$.
An important observation is that {\it the measurement relation is the
same as non-orthogonality}\footnote{The non-orthogonality relation has indeed been used to
 introduce an accessibility relation in the orthoframe semantics within quantum logic \cite{G, Goldblatt}.
}: $s\to t$ iff $s\not\perp t$.

\bigskip\par\noindent\textbf{Quantum actions}
\par\noindent
 A {\it quantum action} is any relation
$R\subseteq \Sigma\times \Sigma$ that can be written as an arbitrary\footnote{That is, possibly infinite.}
union $R= \bigcup_i F_i$ of linear maps $F_i:\Sigma\to \Sigma$. The family of quantum actions forms
a {\it complete lattice} (with inclusion), having set-theoretic {\it union} $R\cup R'$ as supremum. Notice also that
this family is closed under {\it relational composition}
$$R; R' := \{ (s,t)\in \Sigma\times \Sigma: \exists w\in \Sigma (s, w) \in R, (w, t) \in R'\}$$
and {iteration}
$R^* := \bigcup_{k\geq 0} R^n$
(where $R^n= R; R; \cdots R$ is a composition of $n$ terms).
 Quantum actions are a {\it relational (input-output) representation of quantum programs}. Indeed, in our dynamic logic we will interpret (the dynamic modalities for)
quantum programs as (weakest preconditions of) quantum actions.

\bigskip
\par\noindent
\textbf{Weakest precondition, image, strongest post-condition and measurement modalities}
\par\noindent
For any property $T\subseteq \Sigma$ and any quantum action $R\subseteq \Sigma\times \Sigma$, let $$[R] T := \{s\in \Sigma : \forall t\in \Sigma (sR t\Rightarrow t\in T)\} \mbox{ and }
\langle R \rangle T :=  \Sigma \backslash ([R] (\Sigma \backslash
T)).$$
Similarly, put $$R(T):=\{s\in \Sigma: \exists t\in T \mbox{ such that } tRs\}.$$
We also put $R[T]:=\overline{ R(T)}$ for the biorthogonal closure of the image.
Finally, put
$\Box T := \{s \in \Sigma : \forall t (s \to t \Rightarrow t \in T)\} $ and $\, \Diamond T := \Sigma \backslash (\Box (\Sigma \backslash T))$.

\medskip

Observe that $[R]T$ expresses the {\it weakest precondition} for the
`program' $R$ and post-condition $T$. In particular, $[S?]T$  expresses the weakest precondition ensuring the satisfaction of property $T$ in any state after the system passes a quantum test of property $S$. Similarly, $\langle S? \rangle T$ means that one can
perform a quantum test of property $S$ on the current state, ending up
in a state having property $T$. $R(T)$ is the {\it image} of $T$ via $R$,
which is in fact the {\it strongest property (among all properties in ${\mathcal P}(\Sigma\times\Sigma)$ ) ensured to hold after applying program $R$ if a precondition $T$ holds at the input-state}. This is the `strongest postcondition' in an absolute sense. However, the {\it strongest testable postcondition} (ensured to hold after running $R$ if precondition $T$ holds at the input state) is given by $R[T]$. $\Box T$ means that property $T$ will hold after {\it any} measurement (quantum test) performed on the current state. Finally, $\Diamond T$ means that property $T$ is {\it potentially satisfied}, in the sense that one can do some quantum test to reach a state with property $T$.

\blm
{\rm For every property $S\subseteq \Sigma$, we have
$S^{\perp}=[S?]\emptyset=\Sigma\setminus \Diamond S$ and
$\overline{S}= \Box\Diamond S$.
}\elm

\bpr
{\rm For every property $S\subseteq \Sigma$, if $T\in {\cal L}$ (in other words, is testable), then $\Box S, S^{\perp}, [S?]T\in {\cal L}$ (are testable), and more generally $[R] T\in {\cal L}$, for every quantum relation $R$. For every state $s\in \Sigma$, we have $\{s\}\in {\cal L}$, that is, `{\it states are testable}'.
}\epr

\bpr $\, \,$
{\rm A property $S \subseteq \Sigma$ is testable if and only if any of the
following equivalent conditions hold:
\begin{itemize}
\item  $S = \overline{S}$;
\item  $\exists T \in \Sigma \mbox{ \, such \, that } S = T^{\perp}$;
\item  $\exists T \in \Sigma \mbox{\, such \, that } S = \Box T$.
\end{itemize}
}\epr

\bigskip
\noindent\textbf{Quantum joins}
\par\noindent
The family ${\cal L}$ of testable properties is a {\it complete lattice} with
respect to inclusion, having as its meet set-intersection $S\cap T$,
and as its join the biorthogonal closure of set-union $S\sqcup T:= \overline{S\cup T}$,
called the {\it quantum join} of $S$ and $T$. For any arbitrary property
$S\subseteq \Sigma$, we have $\overline{S}=\bigsqcup\{\{s\}: s\in S\}=
\bigcap\{T\in {\cal L}: S\subseteq T\}$, so the biorthogonal closure of
$S$ is the strongest testable property implied by (the property) $S$.

\begin{Th}
{\rm The following properties hold in every quantum frame $\Sigma=\Sigma({\cal H})$:

\begin{enumerate}
\item {\it Partial functionality}
\\
If $s \stackrel{S?}{\to}t$ and $s \stackrel{S?}{\to} v$ then $t = v$.
\item {\it Trivial tests}
\\
$\stackrel{\emptyset ?}{\to} = \emptyset$ and
$\stackrel{\Sigma?}{\to} = \Delta_{\Sigma}$, where
$\Delta_{\Sigma} = \{(s,s) : s \in \Sigma\}$ is the identity relation on $\Sigma \times \Sigma$.
\item {\it Atomicity}
\\
 States are testable, that is,
$\{s\}\in {\cal L}$. This is equivalent to requiring that `states can be distinguished by
tests', that is, \\ $\mbox{ if } s\not= t \mbox{ then } \exists P\in{\cal L}:\, \, \,   s\perp P, \,  t\not\perp
P$.
\item {\it Adequacy}
\\
Testing a true property does not change the state: $\mbox{ if } s \in P \mbox{ then } s \stackrel{P?}{\to} s$
\item {\it Repeatability}
\\
Any testable property holds after it has been successfully tested:\\
$\mbox{ if } s \stackrel{P?}{\to} t \mbox{ then } t \in P$
\item {\it Compatibility}
\\
If $S, T\in {\cal L}$ are testable and $S?;T?$ = $T?;S?$ then $S?;T? = (S\cap T)?$.
\item {\it Self-Adjointness}
\\
If $s \stackrel{P?}{\to} w {\to} t$, then there exists some element $ v \in \Sigma$ such that $t \stackrel{P?}{\to} v {\to} s$.
\item {\it Proper superposition}
\\
Every two states of a quantum system can be properly superposed into a new state:
$\forall s,t \in \Sigma \,\, \exists w \in \Sigma \, \, s {\to} w {\to} t$.
\item {\it Unitary Reversibility and Totality}
\\
Basic unitary evolutions are {\it total bijective functions, having as adjoint
their inverse}:
$$U; U^{\dagger}=U^{\dagger};U =id$$
where $id$ is the identity map.
\item {\it Orthogonality preservation}
\\
Basic unitary evolutions preserve (non) orthogonality:
Let $s,t,s',t'  \in \Sigma$ be such that $ s \stackrel{U}{\to} s'$ and
$ t \stackrel{U}{\to} t'$; then, $ s \to t \mbox{ iff } {s' \to t'}$.
\end{enumerate}
}
\Eth

{\it Proofs}:
\begin{enumerate}
\item {\it Partial functionality} follows from the fact that projectors correspond
to partially defined maps in ${\cal H}$.
\item {\it Trivial tests} follows from the fact that projecting on the empty space yields the empty space and that projecting on the total space does not change anything.
\item {\it Atomicity} follows from the fact that states are nothing but one-dimensional closed linear subspaces, that is, atoms of the lattice of all closed linear subspaces.
\item {\it Adequacy} follows from the fact that for every $x \in W$ we have that $P_W(x) = x$.
\item {\it Repeatability} follows from the fact that $P_W(x) \in W$ for every $x \in {\cal H}$.
\item {\it Compatibility} follows from the fact that if two projectors commute, that is, $P_W \circ P_V = P_V \circ P_W$, then $P_W \circ P_V = P_{W \cap V}$.
\item {\it Self-adjointness} follows from the more general
Adjointness theorem stated below, together with the fact that projectors are self-adjoint (that is, $S?^{\dagger}=S?$).
\item {\it Proper superpositions} can be proved by cases:
\par\noindent
If $s \not \perp t$, that is, let $s \to t$, then $w = s \Rightarrow s \to s \to t$.
\par\noindent
If $s \perp t$, that is, let $s \not \to t$ then let
$s = \overline{x}, t = \overline{y}$ with $ x, y \in {\cal H}$.  Take the superposition $x + y \in {\cal H}$ of $x $ and $ y $ and note that  $x + y \not = 0$ (since from $x + y = 0 \Rightarrow x = - y \Rightarrow s = t$ which contradicts $s \not \perp t$).  Next observe that $x \not \perp( x + y)$ (Indeed, suppose $x  \perp  (x + y)$ then
$\langle x \mid  x + y \rangle = 0$ and then $\langle x \mid x \rangle + \langle x \mid y \rangle = 0$; but $x \perp y$ implies $\langle x \mid x \rangle = 0$. So from $\langle x \mid x \rangle = 0$ follows that $x = 0$, which yields a contradiction). Similarly, we get $y \not \perp (x + y)$.
\par\noindent
Conditions 9 and 10 are immediate consequences of the definition of a unitary operator.
\end{enumerate}

\smallskip

Note that, as a consequence of the `Proper superpositions' property, {\it the double-box modality $\Box\Box$ coincides with
the universal modality}, that is, $\Box\Box S\not=\emptyset$ iff $S=\Sigma$.

\par\noindent
\bth {\bf (Adjointness)}. $\, \,$
{\rm Let $F$ be a quantum map and let $s,w,t \in \Sigma$ be
states: $\, \, \, \mbox{ If }s \stackrel{F}{\to} w {\to} t$ then there
exists some state $v \in \Sigma$ such that
$t \stackrel{F^{\dagger}} {\to} v {\to} s$.
}\Eth
{\it Proof}. \,
To prove this theorem we use the definition of adjointness in a Hilbert space:
$ \langle F x \mid y \rangle = \langle x \mid F^{\dagger} y \rangle$.
From this, we get the equivalence $\langle Fx \mid y \rangle = 0$ iff $\langle x, F^{\dagger} y \rangle = 0$; or, put another way, $Fx  \perp y $ iff $x  \perp F^{\dagger}y $.
Taking the negation of both sides and using the fact that
the measurement relation $s{\to }t$ is the same as non-orthogonality
$s\not\perp t$, we obtain the equivalence:
$\exists w (\overline{x} \stackrel{F}{\to} \overline{w} \to
\overline{y})$ iff
$\exists v ( \overline{y} \stackrel{F^{\dagger}}{\to} \overline{v} \to
\overline{x})$.

\medskip
\par

This proves the adjointness property. As a
consequence, we have the following corolaries.
\smallskip\par\noindent
\bcl{\rm
 For every property $P \subseteq \Sigma$ and every linear map $F$ we have $$P \subseteq [F] \Box \langle F^{\dagger} \rangle \Diamond P.$$
}\ecl

\bcl{\rm
If $F$ is a quantum map,
$$F^{\dagger}(s)= \left( [F] s^{\perp}\right)^{\perp}.$$
}\ecl

{\it Proof}. \,
Using the fact that the negation of the measurement accessibility relation $\to$ is the orthogonality
relation $\perp$, we immediately obtain from the above Adjointness theorem that
$$s \perp F^{\dagger}(t) \mbox{ iff } t\perp F(s) ,$$
that is,
$$s\in (F^{\dagger} (t))^{\perp} \mbox{ iff } F(s) \in t^{\perp}.$$
From this, we get $(F^{\dagger}(t))^{\perp}= [F] t^{\perp}$. Since $F^{\dagger}$ is a {\it map},
$F^{\dagger}(t)$ is a (single) state, so it is a {\it testable} property. Hence, we have
$F^{\dagger} (t) = (F^{\dagger}(t))^{\perp\perp} = ([F] t^{\perp})^{\perp}$.

\medskip

This result leads us to the following natural generalisation of the notion of {\it adjoint}
to {\it all quantum actions}.

\bigskip\par
\noindent\textbf{Adjoint of a quantum action}
\par\noindent
For every quantum action $R\subseteq \Sigma\times \Sigma$, we define
a relation $R^{\dagger}\subseteq \Sigma\times\Sigma$ by
$$s R^{\dagger} t \, \mbox{ iff } t\perp [R]s^{\perp},$$
or, put another way,
$$R^{\dagger} (s) = \left( [R] s^{\perp} \right)^{\perp}.$$

\smallskip

\bpr
{\rm For all quantum actions $R,Z\subseteq \Sigma\times \Sigma$, states $s, t\in \Sigma$ and properties $S\subseteq \Sigma$, we have the following:
\begin{enumerate}
\item $R^{\dagger}$ is a quantum action.
\item If $R=F$ is a (quantum, that is, linear) map\footnote{We identify a map $F:\Sigma\to \Sigma$ with its
graph $F\subseteq \Sigma\times\Sigma$, that is, quantum maps are special cases of quantum relations, which happen to be
be partial functions. So
$R=F$ means that the two sides are equal, as relations.}
 then the relational adjoint $R^{\dagger}$ coincides with the Hermitian
adjoint $F^{\dagger}$ (of $F$ as linear map).
\item $s \perp R^{\dagger}(t)$ iff $t\perp R(s)$ .
\item $(R; Z)^{\dagger} = Z^{\dagger}; R^{\dagger}$.
\item $(R\cup Z)^{\dagger} = R^{\dagger} \sqcup Z^{\dagger}$.
\item $R[S] = ( [R^{\dagger}] S^{\perp} )^{\perp}$.
\end{enumerate}

}
\epr


\subsection{Compound-system quantum frames}

In this subsection we extend the quantum frame presented above
for single systems into a quantum frame for compound systems.  Let ${H}$
be a Hilbert space of dimension $2$ with basis $\{ \mid 0 \rangle,
\mid 1 \rangle \}$.  We fix a natural number $n\geq 2$ (although later
we will restrict consideration to the case $n\geq 4$), and put $N=\{1,2,\ldots , n\}$. Our global state space will be denoted as before by ${\cal H}$, but now we assume it is an $n$-qubit state, that is, we put ${\cal H} = {\cal H}_{n} := H^{\otimes n} = H \otimes H \otimes ... \otimes H$ ($n$ times) for the tensor product of $n$ copies of $H$.
A {\it $n$-qubit quantum frame} will be the
quantum frame $\Sigma :=\Sigma({\cal H})$ associated (as in the previous section) to the Hilbert space ${\cal H}$.

\bigskip\par\noindent
{\bf Notation}
\par\noindent
In fact, we consider all the $n$ copies of ${H}$ as
distinct (although isomorphic) and use $H^{(i)}$ to denote the $i$-th
component of the tensor $H^{\otimes n}$. Also, for any set of indices $I\subseteq N$, we put
${\cal H}_I=H^{\bigotimes I}= \bigotimes_{i\in I}H^{(i)}$. Note that we have ${\cal H}_N={\cal H}_n= {\cal H}$. We use $\epsilon_i: H\to H^{(i)}$ to denote the canonical isomorphism
between ${\cal H}$ and ${\cal H}^{(i)}$. This notation can be extended
to sets $I\subseteq N$ of indices of length $|I|=k$, by putting
$\epsilon_I: H^{\otimes k}\to {\cal H}_I$ to be the canonical
isomorphism between these spaces.
Similarly, for each set
$I\subseteq N$, we use $\mu_I: {\cal H}_I\otimes {\cal
H}_{N\setminus I} \to {\cal H}$ to denote the canonical isomorphism between
these two spaces. For any
vector $\mid x \rangle\in H$, we use $\mid x\rangle^{\bigotimes
I}=\bigotimes_{i\in I} \mid x\rangle$ to denote the corresponding vector
in ${\cal H}_I$ (obtained by tensoring $|I|$ copies of $\mid x\rangle$ ).
Given a set $I\subseteq N$, we say that a state $s\in
\Sigma({\cal H})$ {\it has its $I$-qubits in state} $s'\in
\Sigma({\cal H}_I)$, and write $s_I=s'$, if there exist vectors
$\psi\in s$, $\psi'\in {\cal H}_I$ and $\psi''\in {\cal H}_{N\setminus
I}$ such that $\psi= \mu_I(\psi'\otimes \psi'')$. Note that the state
$s_I$, {\it if it exists, is unique} (having the above property).
We say that the state $s$ is {\it $I$-separated} iff $s_I$ exists. In this case,
$s_I$ is called the {\it ($I$-)local component} (or {\it local state}) of $s$.
In particular, when
$I=\{i\}$, the local component $s_i\in {\cal H}_{\{i\}}=H^{(i)}$
is called {\it the $i$-th coordinate} of
the state $s$.

We will further use $\mid + \rangle$ to denote the vector $\mid 0 \rangle + \mid 1 \rangle$, and, similarly $\mid - \rangle$ to denote
$\mid 0 \rangle - \mid 1 \rangle$.  For the states generated by the vectors in a two dimensional Hilbert space we introduce the following abbreviations:
$+ := \overline{\mid + \rangle}$, $- := \overline{\mid - \rangle}$ , $
0 := \Odiraco$ , $ 1 := \Odirac$.
 In order to refer to the state corresponding to a pair of qubits, we similarly delete the Dirac notation, for example, $00 := \overline{\mid 00 \rangle} = \overline{\mid 0 \rangle \otimes \mid 0 \rangle}$.
The Bell states will be abbreviated as follows:
$\beta_{00} := \overline{\mid 00\rangle +\mid 11\rangle }$ , $\beta_{01} := \overline{\mid 01 \rangle + \mid 10 \rangle }$, $ \beta_{10} := \overline{\mid 00 \rangle -
\mid 11 \rangle}$ , $\beta_{11} = \overline{\mid 01 \rangle - \mid 10 \rangle}$
and $\gamma := \overline{\mid 00\rangle + \mid 01\rangle + \mid 11\rangle +\mid 10\rangle}$.

\bigskip

The following two results are well-known.
\par\noindent
\bpr{\rm
 Let ${ H}^{(i)} $ and $ { H}^{(j)}$ be two Hilbert spaces.  There exists a bijective correspondence $\psi$ between the linear maps $F : {H}^{(i)} \to { H}^{(j)}$ and the states of ${ H}^{(i)} \otimes { H}^{(j)}$.  For fixed bases $\{\epsilon_{\alpha}^{(i)}\}_{\alpha}$ and $\{\epsilon_{\beta}^{(j)}\}_{\beta}$  of these spaces, the correspondence $\psi$ is maps the linear function $F$, given by
$ F(\mid x \rangle) = \Sigma_{\alpha \beta} \, m_{\alpha\beta} \,  \langle \epsilon_{\alpha}^{(i)} \mid x \rangle . \epsilon_{\beta}^{(j)}$ for all $\mid x \rangle \in H^{(i)}$, to the state $\psi(F) = \Sigma_{\alpha \beta} \, m_{\alpha \beta} \, . \epsilon_{\alpha}^{(i)} \otimes \epsilon_{\beta}^{(j)}$.
}\epr

\bpr
{\rm Let ${\cal H} = H^{\otimes n}$ and let $W = \{x \, \otimes \mid 0
\rangle^{\otimes (n-1)} : x \in H\}$ be given. Any linear map $F:
{\cal H} \to {\cal H}$ induces a linear map ${F_{(1)}: { H} \to { H}}$ in a
canonical manner: it is defined as the unique map on ${ H}$ satisfying
$F_{(1)}(x) =$ ${P_W \circ F(x \, \otimes \mid 0 \rangle^{\otimes (n-1)})}$.
Conversely, any linear map $G: H \to H$ can be represented as $G = F_{(1)}$ for some linear map $F: {\cal H} \to {\cal H}$.
}\epr

\medskip
\par\noindent
{\bf Notation}
\par\noindent
The above results allow us to specify a compound
state in $H^{(i)} \otimes H^{(j)}$ via some linear map $F$ on ${\cal
H}$. Indeed, if  $F : {\cal H} \to {\cal H}$ is any such linear map,
let $F_{(1)}: H\to H$ be the map in the above proposition; this induces a
corresponding map $F_{(1)}^{(ij)}: H^{(i)} \to H^{(j)}$, by
putting $F_{(1)}^{(ij)} := \epsilon_j \circ F_{(1)} \circ \epsilon_i^{-1}$,
where  $\epsilon_i$ is the canonical isomorphism introduced above (between
${H}$ and the $i$-th component ${H}^{(i)}$ of  ${H}^{\otimes n}$).
Then we use $\overline{F}_{(ij)}$ to denote the state
$$\overline{F}_{(ij)} := \overline{\psi(F_{(1)}^{(ij)})}$$
given by the above mentioned bijective
correspondence $\psi$ between $H^{(i)} \to H^{(j)}$ and $H^{(i)}
\otimes H^{(j)}$. The following result is also known from the
literature.

\medskip
\bpr{\rm
Let  $F : {\cal H} \to {\cal H}$ be a linear map. Then the state
$\overline{F}_{(ij)}$ is `entangled according to $F$'; that is, if $F_{(1)}(\mid x\rangle)=\mid y\rangle$ and the state of a
2-qubit system is $\overline{F}_{(ij)}\in H^{(i)}\otimes H^{(j)}$, then any measurement
of qubit $i$ resulting in a state $x_i$ collapses the qubit $j$ to
state $y_j$.
}\epr

In our axiomatic proof system, we will take (a syntactic counterpart of) this result as our
central axiom, the `Entanglement Axiom'.

\medskip
\par\noindent
{\bf Notation}
\par\noindent
The notation $\overline{F}_{(ij)}$ can be further extended to
define a property (set of states) $\overline{F}_{ij}\subseteq \Sigma= \Sigma({\cal H})$, by defining it as {\it the set of all
states having the $\{i,j\}$-qubits in the state
$\overline{F}_{(ij)}$ }:
\begin{eqnarray*}
\overline{F}_{ij} \, & = & \, \{s\in \Sigma:
s_{\{i,j\}}=\overline{F}_{(ij)}\} \\
& = & \, \{ \overline{\mu_{\{i,j\}}(\psi\otimes\psi')} :
\psi\in \overline{F}_{(ij)},
\psi' \in
{\cal H}_{N\setminus \{i, j\}} \}\subseteq \Sigma
\end{eqnarray*}

\noindent where $\mu_{\{i,j\}}$ is, as above, the canonical isomorphism between
${\cal H}_{\{i,j\}}\otimes {\cal H}_{N\setminus \{i,j\}}$. In other
words,
$\overline{F}_{ij}$ is simply the property of an $n$-qubit compound
state of having its $i$-th and $j$-th qubits (separated from the
others, and) in a state that is
`entangled according to $F$'.

\bigskip
\par\noindent
{\bf Local properties and separation}
\par\noindent
 Given a set $I\subseteq N$, a property
$S\subseteq \Sigma$
is {\it $I$-local } if it corresponds to a property of the
subsystem
formed by the qubits in $I$; in other words, if there exists some
property $S'\subseteq \Sigma({\cal H}_I)$ such that:
$$
S =\{s\in \Sigma: s_I \in S'\} $$
or, more explicitly,
$S=\{\overline{\mu_I(\phi\otimes \psi)}: \overline{\phi}\in S', \psi\in
{\cal H}_{N\setminus I}\}$.
An {\it example} is the property $\overline{F}_{ij}$, which is
$\{i,j\}$-local.
For any $I \subseteq N$, the family of $I$-local properties forms a {\it complete lattice} (with inclusion)
in which the join is given by {\it union} $S\cup T$, the
{\it atoms} correspond to {\it local states}, and the {\it greatest element} is the
property
$$\top_I^{\Sigma} := \{s\in\Sigma: s \mbox{ is } I-\mbox{separated}\}=
\bigcup \{ S\subseteq \Sigma: S \mbox{ is } I-\mbox{local} \}$$
that defines {\it separation}: a state $s$ is $I$-separated iff $s\in \top_I^{\Sigma}$. But note that the family of $I$-local properties
is {\it not closed under complementation}.

\bigskip
\par\noindent
{\bf Local Maps}
\par\noindent
Given $I\subseteq N$, a linear map $F: {\cal H}\to {\cal
H}$
is {$I$-local} if it `affects only the qubits in $I$'; in other
words, if there exists a map $G:{\cal H}_I\to {\cal H}_I$ such that
$$F\circ\mu_I \, (\phi \otimes\psi) \, =\, \mu_I \,( G(\phi) \otimes \psi)$$
A map $F:\Sigma\to \Sigma$ is {\it $I$-local} if it is the map induced
on $\Sigma$ by an $I$-local linear map on ${\cal H}$. {\it Examples} are:
all the tests $S_I ?$ of testable $I$-local properties $S_I$; logic gates that affect
only the qubits in $I$, that is, (maps on $\Sigma$ induced by) unitary
transformations $U_I: {\cal H}\to {\cal H}$ such that for all $\psi \in {\cal H}_{I},
\psi'\in {\cal H}_{N\setminus I}$, we have $U_I\circ \mu_I(\psi\otimes \psi')=
\mu_I ( U(\psi)\otimes \psi')$, for some $U:{\cal H}_I\to {\cal H}_I$. The
family
of $I$-local maps is closed under composition.

\medskip
\par\noindent
{\bf Local actions}
\par\noindent
An {\it $I$-local action} is a quantum action $R\subseteq \Sigma\times\Sigma$ that
that can be written as an arbitrary\footnote{That is, possibly infinite} union of $I$-local maps.
The family of $I$-local actions forms a {\it complete lattice} (with inclusion), in which
the join is given by {\it union} $R\cup R'$, and the {\it greatest element} is the action
$$\top_I^{\Sigma\times \Sigma} := \bigcup \{ F:\Sigma \to \Sigma: F \mbox{ is an } I-\mbox{local map } \}$$

\medskip

\par\noindent
\blm {\bf (Teleportation property).}
{\rm If $s$ is an $i$-separated state having its $i$-th qubit $s_i$ in the state
$x\in H$, then after doing two successive bipartite measurements $\overline{G_{jk}}?$ followed
by $\overline{F_{ij}}?$,
the $k$-th qubit ($k$-th component of) the output-state is
$$ \left( \overline{F_{ij}}? \circ \overline{G_{jk}}? (s)\right)_k = G_{(1)}\circ F_{(1)} (x)$$
}\elm

\medskip
\par\noindent
\blm {\bf (Entanglement Composition Lemma).}
{\rm The main lemma in \cite{Coecke2} states (in our notation) that, given
a quadruple of {\it distinct} indices $i, j, k, l$,  and leting
$F, G, H, U, V: H\to H$ be single-qubit linear
maps (that is, $1$-local transformations), we have
$$\overline{G_{jk}}? \circ V_k \circ U_j\, ( \overline{F}_{ij}\cap
\overline{H}_{kl})\, \subseteq \, \overline{(H\circ U^{\dagger}\circ G \circ
V \circ F)}_{il}.$$
}\elm

\smallskip
\cite{Coecke2} and \cite{AbramskiCoecke} use these last two lemmas as the main tool in explaining teleportation,
quantum gate teleportation
and many other quantum protocols. We will use this work in our logical
treatment of such protocols by formally proving (syntactic correspondents of) these lemmas in
our axiomatic proof system and then using them to analyse teleportation and quantum secret sharing.

Observe that in the above Lemma 3, the order in which the operations
$U_j$ and $V_k$ are applied is in fact {\it irrelevant}. This is a consequence of
the following important property of local transformations.

\medskip
\par\noindent
\bpr {\bf (Compatibility of local transformations affecting different
sets of qubits).}
{\rm If $I\cap J=\emptyset$, $F$ is an $I$-local map and
$G$ is a $J$-local map, we have
$$F \circ G = G \circ F.$$
}\epr

\medskip

Another important property of local maps (on {\it states}) is given by the following proposition.

\par\noindent
\bpr {\bf (Agreement Property).}
{\rm Let $F_I, G_I: \Sigma\to \Sigma$
be two $I$-local maps
on states, having the same domain\footnote{The domain of a map is
defined by $dom(F)=\{s\in \Sigma: F(s) \mbox{ is defined } \}$. If $F'$
is the corresponding linear map on ${\cal H}$, this means that
$dom(F)=\{\overline{\psi}: F'(\psi)\not=0\}$.}:
$dom(F)=dom(G)$.
Then their output-states agree on all non-$I$ qubits, that is, for all $s\in \Sigma$,
$$F(s)_{N\setminus I} = G(s)_{N\setminus I}$$
whenever both sides of the identity {\it exist} (that is, whenever both
$F(s)$ and $G(s)$ are {\it $I$-separated}.)
}\epr

\newpage
\par\noindent{\bf Dynamic characterizations of main unitary transformations}
\smallskip\par\noindent
It is well-known that a linear operator on a vector space in a given
Hilbert space is {\it uniquely determined} by the values it takes on
the vectors of an (orthonormal) basis.  An important observation is
that this fact is no longer `literally true' when we move to
`states' as one-dimensional subspaces instead of vectors. The reason
is that `phase'-aspects (or, in particular, the signs `$+$' and
`$-$') are not `state' properties in our setting.  In other words,
two vectors that differ only in phase, that is, $x = \lambda y$ where $\lambda$ is a complex number
with $\mid \lambda \mid = 1$, belong to the same subspaces, so they correspond to the
same state $\overline{x} = \overline{y}$.
\medskip

\bex {\bf (Counterexample).} $\, \,$
\rm{
Consider a 2 dimensional Hilbert space in which we use $\mid 0 \rangle$
and $\mid 1 \rangle$ to denote the basis
vectors. A transformation $I$ is given by $I(\alpha {\mid
0 \rangle} + \beta {\mid 1 \rangle}) = \alpha \mid 0
\rangle + \beta \mid 1 \rangle$; and a transformation $J$ is given by $J(\alpha {\mid 0 \rangle} + \beta {\mid 1 \rangle}) = \alpha \mid 0 \rangle - \beta \mid 1 \rangle$.  Although $I $ and $J$ induce different operators on states, these operators map the basis states to the same images:
$$I(0) = \overline{I(\mid 0 \rangle)} = 0 = \overline{J(\mid 0 \rangle)} = J(0),$$
$$I(1) = \overline{I(\dirac)} = 1 = \overline{- \dirac} = \overline{J(\dirac)} = J(1).$$
But of course we do distinguish the subspaces generated by different superpositions:
$$I(+) = \overline{\diraco + \dirac}  = + \not = - = \overline{\diraco - \dirac} = J(+).$$
}
\eex
\medskip

\bpr{\rm
A linear operator on the state space $\Sigma({\cal H}_1)$ of a 2 dimensional Hilbert space is
uniquely determined by its images on the states: $\Odiraco, \Odirac, \overline{\mid + \rangle}$.
}\epr
\medskip

\bcl{ \rm
A linear operator on the state space $\Sigma({\cal H}_n)$ of the space ${\cal H}_n$ is uniquely determined by its
images on the states:
$$\{\overline{ \mid x \rangle_1 \otimes ... \otimes \mid x \rangle_n} : \mid x \rangle_i \in \{ \mid 1\rangle_i, \mid 0\rangle_i, \mid +\rangle_i \}\}$$
}\ecl

\bigskip
\par\noindent
In the definition of a quantum frame given above, we introduced the set ${\cal U}$
as the set of unitary transformations for single systems.
For compound systems the set ${\cal U}$ will be extended with the kind of operators
that are active on compound systems.  Following the quantum computation literature, we take
${\cal U} = \{ X, Z, H, CNOT, ...\}$ where $X, Z$ and $H$ are defined by the following table:
\bigskip\par\noindent
\centerline{\begin{tabular}{r||r|r|r}
 & 0 & 1 & + \\
 \hline \hline
 X& 1 & 0 & + \\
 \hline
 Z & 0 & 1 & - \\
 \hline
 H & + & - & 0
 \end{tabular}}
\bigskip\par\noindent
The transformation $CNOT$ is given by the table:
\bigskip\par\noindent
\begin{tabular}{r||r|r|r|r|r|r|r|r|r}
&        $00$&  $01$&  $0+$& $11$& $10$& $1+$ &$+0$         &$ +1$        &$++ $\\
\hline \hline
$CNOT $& $00$& $01$&   $0+$& $10$& $11$& $1+$ &$\beta_{00}$ &$\beta_{01}$ &$ \gamma $
 \end{tabular}

\section{The logic $LQP$}

\subsection{Syntax of $LQP$}
\smallskip\par\noindent
To build up the language of $LQP$, we are given the following: a natural number $n$, for which we put $N=\{1, 2, \ldots, n\}$; a set ${\cal Q}$ of {\it propositional variables}; a set ${\cal C}$
of {\it propositional
constants}; and a set ${\cal U}$ of program
constants, denoting {\it basic programs}, to be interpreted as {\it quantum gates}.
Each program constant $U\in {\cal U}$ {\it comes together with an index} $I$,
which is a sequence
of distinct indices in $N$; the index gives us the set of qubits on which the quantum gate $U$
is active - when we want to make explicit the index, we write, for example,
$U_I$ for an $I$-local quantum gate. In particular, for every $i,j\leq n$, we are given some special
program constants $CNOT_{ij}, X_i, H_i, Z_i,\ldots\in {\cal U}$. Similarly, we are given two special
propositional
constants $1, +\in {\cal C}$, the first denoting the separated state $\mid 1\rangle^{\otimes n}=\mid 1\rangle\otimes
\mid1 \rangle \cdots \otimes \mid1\rangle$ and the second denoting the separated state
$\mid+\rangle^{\otimes n}= \mid+\rangle\otimes
\mid+ \rangle \cdots \otimes \mid+\rangle$.
The syntax of $LQP$ is an extension of the classical syntax for $PDL$,
with a set of propositional {\it formulas} and a set of {\it programs}, defined by mutual induction:
\beqa
\begin{array}{lllllllllllllll}
\varphi & ::=   &\top_I  & \mid & p    & \mid  & c & \mid & \neg \varphi  &  \mid & \varphi \wedge \varphi & \mid & [\pi] \varphi   \\
\pi & ::=        & \top_I & \mid & \varphi? &  \mid & U & \mid & \pi^{\dagger} & \mid  & \pi \cup \pi           & \mid & \pi ; \pi
\end{array}
\eeqa
Here, we take $I$ to denote sequences of distinct indices in $N=\{1,2,\ldots, n\}$.
The sentence $\top_I$ expresses {\it $I$-separation}: it is true iff the qubits in $I$ form a separated subsystem.
So $\top_I$ denotes the greatest element $\top_I^{\Sigma}$ of the lattice of $I$-local properties.
In particular, the sentence $\top_N$ denotes the `always true' proposition ({\it verum}, usually denoted $\top$), tat is, the `top' of the lattice of all properties.\footnote{Note also the distinction between the constant $1_i$ (characterising the qubit
$\mid 1\rangle_I$) and the constant $\top_i$ (denoting the property of being $i$-separated).}
The constructs $\neg\varphi$ and $\varphi\wedge\varphi$ denote classical negation and conjunction, while
the construct given by dynamic modalities $[\pi]\varphi$ denotes {\it the weakest precondition that ensures that property
$\varphi$ will hold after running program $\pi$}.

On the program side: $\top_I$ denotes the {\it trivial $I$-local action} $\top_I^{\Sigma\times\Sigma}$,
which acts on any given $I$-separated state
by keeping the $N\setminus I$ subsystem unchanged, while changing the $I$ subsystem to any randomly picked $I$ system.
In other words, $\top_I$ is the union of all $I$-local actions. The meaning of quantum test $\varphi?$, adjoint $\pi^{\dagger}$,
union $\pi\cup \pi$ and composition $\pi;\pi$ is given by the corresponding operations on quantum actions.

Notice that we did not include {\it iteration} (Kleene star) among our program constructs: this is only
because we do not need it for any of the applications in this paper. Indeed, most quantum programming does not
involve $while$-loops; but (as pointed in our Section 6)
one can of course add iteration to our logic, if needed.

\medskip
\par\noindent
{\bf Abbreviations in $LQP$}
\par\noindent
We can enrich our basic language by introducing various abbreviations. In particular, we define the {\it classical disjunction} and {\it classical implication} in the usual way, that is,
$\varphi \vee \psi := \neg (\neg \varphi \wedge \neg \psi)$,
$\varphi \to \psi := \neg \varphi \vee \psi$. As in classical logic, we can introduce a constant {\it falsum} $\bot_N := \neg \top_N$ for the `always false' sentence (usually denoted
$\bot$). In non-ambiguous contexts, we sometimes skip the subscript $N$, and simply write $\top$ and $\bot$ for $\top_N$ and $\bot_N$. We define the {\it classical dual} of $[\pi] \varphi$ in the usual way as
$\langle \pi \rangle \varphi := \neg [\pi] \neg \varphi$ ;
the {\it measurement modalities} $\Box$ and $\Diamond$ used in the quantum logic literature can be defined in $LQP$ by putting
$\Diamond \varphi : = \langle \varphi ? \rangle \top_N$ and $\Box \varphi
:= \neg \Diamond \neg \varphi$.
The {\it orthocomplement} is defined as $\sim \varphi := \Box \neg \varphi$, or, equivalently, as $\sim \varphi := [\varphi?]\bot_N$. Using the orthocomplement, we define
a binary operation for {\it quantum join}
$\varphi \sqcup \psi := \sim (\sim \varphi \wedge \sim \psi)$.
This expresses {\it superpositions}: $\varphi\sqcup \psi$ is true at
any state which is a superposition of states satisfying
$\varphi$ or $\psi$.

We also introduce some notions and notations for programs: we
call a program $\pi$ {\it deterministic} if $\pi$ is constructed
without the use of non-deterministic choice $\cup$ or of the non-deterministic program $\top_I$.
Also, we put
$$flip_{ij} \, \, :=\,\, CNOT_{ij} ; CNOT_{ji}; CNOT_{ij}$$
for the program that (given any $\{i,j\}$-separated input state)
permutes the $i^{th}$ and the $j^{th}$ components. Finally, we put
$$id \, \, : = \,\, \top?$$
for the {\it identity} map.

\bigskip
\par\noindent
\textbf{Order, equivalence, orthogonality, $I$-equivalence, testability, locality, separation}
\par\noindent
We can
internalize the {\it logical equivalence, being weaker than, and $I$-equivalence} relations between formulas, the {\it locality} and {\it testability}, and the notion of
{\it $I$-component} by defining
the following formulas:

\begin{eqnarray*}
\begin{array}{lllllll}
\varphi \leq \psi \, \,  & := & \,\, \Box\Box(\varphi \rightarrow \psi) \\
\varphi = \psi \, \, & := & \,\, \Box\Box(\varphi \leftrightarrow \psi) \\
\varphi \perp \psi \, \, & := & \,\, \varphi \leq \sim \psi \\
T(\varphi) \,\, & := & \,\, \sim\sim\varphi \leq \varphi \\
\varphi_I \, \, & := & \,\, \top_I\wedge \langle\top_{N\setminus I}\rangle\varphi \\
\varphi =_I \psi \, \, & := & \, \, \varphi \leq \top_I \, \wedge \psi\leq \top_I \, \wedge\, \varphi_I= \psi_I\\
I(\varphi) \, \, & := & \, \, \varphi=\varphi_{I}.
\end{array}
\end{eqnarray*}
Recall from Section 2.1 that the double-box modality coincides with
the universal modality: so, indeed, $\varphi\leq \psi$ means that $\varphi$ is {\it logically weaker} than
$\psi$, while $\varphi=\psi$ means the formulas are {\it equivalent}.
We read $T(\varphi)$ as saying that `$\varphi$ is testable', and
$I(\varphi)$ as `$\varphi$ is $I$-local'. We read $\varphi_I$ as `the $I$-component of $\varphi$':
a state satisfies this sentence iff (it is $I$-separated and) its $I$-subsystem is (a subsystem of some
state) satisfying $\varphi$. For $I=\{i\}$, we write $\varphi_i:= \varphi_I$.
We read $\varphi =_I \psi$ as `$\varphi$ is $I$-equivalent to $\psi$':
the meaning is that {\it both $\varphi$ and $\psi$ are $I$-separated and have the same $I$-component}.
Finally, we say that $\varphi$ is {\it $I$-separated} iff $\varphi\leq \top_I$.

\medskip

Note that it obviously follows from these definitions that {\it every $I$-component
$\varphi_I$ is $I$-local}.

\smallskip
\par\noindent
\textbf{Special local states}
\par\noindent
We can introduce some more propositional constants (which will denote special local states),
by putting:
$0_i:=\sim 1_i$ and $-_i :=\sim +_i$.

\smallskip
\par\noindent
\textbf{Image and strongest post-condition}
\par\noindent
We define
the {\it strongest testable post-condition $\pi[\varphi]$ ensured by
(applying a program) $\pi$ on (any state satisfying a given precondition)
$\varphi$}, by putting
$$\pi[\varphi] \,\, :=\,\, \sim [\pi^{\dagger}]\sim \varphi$$
If $\varphi$ is assumed to be {\it testable} and $\pi$
is {\it deterministic}, the strongest postcondition $\pi [\varphi]$ coincides with the {\it image} $\pi (\varphi)$ of
$\varphi$ via $\pi$.
The definition of {\it image of a testable property via a program} $\pi(\varphi)$
can be extended to {\it all programs that are finite unions of deterministic programs}, by
putting,
for all {\it testable} formulas $\phi$:
$\pi(\phi)=\pi[\phi]$ if $\pi$ is deterministic, and
$(\pi\cup \pi')(\phi)=\pi(\phi)\vee\pi'(\phi)$ otherwise.

\medskip

Note  the {\it contrast with classical $PDL$}: unlike the classical version, our quantum $PDL$ (as considered
above, that is, {\it without program
converse}\footnote{There also exists a version of $PDL$ {\it with a program converse operator} $\pi^{\smile}$, such that
the accessibility relation for the converse  $\pi^{\smile}$ is defined as the converse of the accessibility
relation for $\pi$. It is obvious that this stronger logic can express the strongest post-condition of a program $\pi$,
using the existential dynamic modalities, since $\pi (\varphi) =\langle \pi^{\smile}\rangle\varphi$.}) has enough
expressive power to {\it define strongest post-conditions (and, in a restrict context, images) using weakest preconditions}!
The reason is that, in some context, the notion of adjoint can replace the notion of converse. But note that converse
itself is {\it not} expressible in our logic. This is a good thing since the converse of a quantum action has no physical meaning (except in
the case of reversible, unitary evolutions), while the adjoint is physically meaningful.

\par\noindent\textbf{Notation}
\par\noindent
For any sequence $I\subseteq N$ of indices and any
vector $\vec{c}=(c(i))_{i\in I}\in \{0,1,+\}^{|I|}$, we set
$$\vec{c}_I := \bigwedge_{i\in I} c(i)_i.$$

\medskip
\par\noindent
\textbf{The unary maps induced by a program}
\par\noindent
In our syntax we want to capture the construction $F_{(1)}$, by which
a linear map $F$ on $H^{\otimes n}$ was used to describe a unary map
$F_{(1)}$
on $H$. For this, we put:
$0_i! \, := \, 0_i? \cup (1_i ? ; X_i)$, and
$0_I ! \, :=\, 0_{i_1}! ; 0_{i_2} ! ; \cdots ; 0_{i_k}!$, where
$I=(i_1, i_2, \ldots, i_k)$. This maps any qubit in $I$ to
$0$. Similarly,
we put; $0_I ? \, := \,  (0_{i_1}\wedge 0_{i_2} \wedge \cdots \wedge
0_{i_k})?$.
Finally, we define
$$\pi_{(i)} := 0_{N\setminus \{i\}}! ; \pi ; 0_{N\setminus \{i\}}?$$
This is the map we need (which encodes a single qubit transformation).
In fact, we shall only use $\pi_{(1)}$ in the rest of this paper.
We also want to consider the $H_i\to H_j$-version of the
transformation $\pi_{(1)}$, so we put
$$\pi_{ij} := flip_{1i}; \pi_{(1)}; flip_{1j}$$

\smallskip
\par\noindent
\textbf{Local programs}
\par\noindent
We would like to isolate {\it local programs}, that is, the ones that `affect only
the qubits in a given set $I \subseteq N$'. For this, we define a
formula $I(\pi)$ meaning `program $\pi$ is $I$-local':
$$I(\pi) \,\, := \,\,  \bigwedge_{\vec{c},\vec{d},\vec{d'}}\left(
\vec{d}_{N\setminus I}\, =_{N\setminus I}\, \pi(\vec{c}_I\wedge \vec{d}_{N\setminus I}) \,  =_I \,  \pi(\vec{c}_I\wedge
\vec{d'}_{N\setminus I})\right)$$
where the conjunction is taken over all $\vec{c}\in\{0,1,+\}^{|I|}$
and all $\vec{d}, \vec{d'}\in \{0,1,+\}^{n - |I|}$.

Note that this definition is a simple formal translation of the semantic clauses that
express the fact that program $\pi$ {\it acts only `locally'} (affecting only the $I$-subsystem, and in a way
that depends only on the $I$-subsystem of the input state) {\it on the states of the form $\vec{c}$}
(with $c\in \{0,1,+\}$). One of our axioms below
(`Determinacy of deterministic programs') means that this clause is enough
to ensure that program $\pi$ acts locally {\it on all ($I$-separated) states}.

\medskip

\par\noindent
\textbf{Entanglement according to $\pi$}
\par\noindent
To describe states
that are `entangled according to $\pi$', we introduce the
following formula
$$\overline{\pi}_{ij} \, := \, \top_{ij}\wedge \bigwedge_{c\in\{0,1,+\}} \left( [c_i?]
(\pi_{ij}(c_i))_j \wedge (\sim c_i \rightarrow
\pi_{ij}(c_i)=\bot)\right).$$
Then, as a consequence, we will have the following obvious validity:
$$c_i? (\overline{\pi}_{ij}) =_j \pi_{ij}(c_i)$$
for every $c_i\in \{0_i,1_i.+_i\}$.

\medskip

Again, note that the identity in this definition is a formal translation of the semantic clause defining
`entanglement according to an action', {\it but only for the particular case of local states of the form $c_i$}
(with $c\in \{1, 0, +\}$). And again, one of our axioms below (the `Entanglement Axiom')
ensures that the above identity holds (not only for the elements $c_i$,
but) {\it for all $i$-local states} (that is, all testable $i$-local properties).

\subsection{Semantics of $LQP$}

An {\it $LQP$-model} is a {\it multi-partite quantum frame} $\Sigma=\Sigma({\cal H})$ based on an $n$-dimensional Hilbert
space ${\cal H}$, together with a {\it valuation function}, mapping each propositional variable $p$ into a set of states
$\M p \M \subseteq \Sigma$ .
We will use the valuation map to give an interpretation $\M
\varphi\M \, \subseteq \Sigma$
to all our formulas, in terms of quantum properties of our multi-partite frame,
that is, sets of states in $\Sigma$. At the same time, we
give an interpretation $\M \pi \M \, \subseteq  \Sigma \times\Sigma$ to all our programs,
in terms of {\it quantum actions}. The two interpretations are defined by {\it mutual recursion}.
\smallskip\par\noindent
{\bf Interpretation of Programs}
\par\noindent
\begin{eqnarray*}
\begin{array}{lllllll}
\M \top_I \M \, & := & \, \top_I^{\Sigma\times\Sigma} & , &
\M \varphi ?\M \, & := & \, \M \varphi \M ? \\
\M U \M \, & := & \, U  & , &
\M \pi^{\dagger} \M  \, & := & \, \M\pi\M ^{\dagger} \\
\M \pi_1 \cup \pi_2 \M  \, & := & \, \M \pi_1 \M \cup \M \pi_2 \M & , &
\M \pi_1;\pi_2 \M \, & := & \,  \M \pi_2 \M ; \M \pi_1 \M
\end{array}
\end{eqnarray*}
The interpretation $\M \pi\M$ allows us to extend the notation
$\stackrel{\pi}{\to}$ to all programs, by putting: $s
\stackrel{\pi}{\to} t$ iff $(s, t)\in\, \M \pi\M$.

\bigskip
\par\noindent
{\bf Interpretation of formulas}
\par\noindent
We extend the valuation
$\M p\M$ from propositional variables to all formulas, by putting for
the others:
\begin{eqnarray*}
\begin{array} {lllllll}
\M 1 \M  \, & = & \, \mid 1\rangle^{\otimes n} &  &
\M + \M  \, & = & \, \mid +\rangle^{\otimes n} \\
\M \varphi \wedge \psi \M  \, & = & \,  \M \varphi \M \,  \cap \, \M \psi \M &  &
\M \neg \varphi \M  \, & = & \, \Sigma \backslash \M \varphi \M \\
\M [\pi]\varphi  \M \, & = & \, [ \,\M\pi\M\, ] \, \M\varphi\M &  &
\M \top_I \M \, & = & \, \top_I^{\Sigma}
\end{array}
\end{eqnarray*}
$$$$
\bpr
{\rm
The interpretation of any testable formula is a testable property. The
interpretation of an $I$-local formula (or $I$-local deterministic program) is
an $I$-local
property (or $I$-local linear map on states).
}
\epr

\blm $$\M\sim \varphi\M=\M\varphi\M^{\perp}$$
$$\M[\varphi?]\psi\M=[\M\varphi\M ?]\M\psi\M$$ $$\M\Box\varphi\M=\Box\M\varphi\M$$
$$\overline{\M \varphi\M }=\M \sim\sim \varphi\M$$
\elm

\bpr
{\rm The following are equivalent, for every formula $\varphi$: \\
\begin{tabular}{ll}
$1.$ & $ \,  \M\varphi\M$  is testable (that is, $T(\varphi)$ is valid).\\
$2. $ & $\, \varphi$  is semantically equivalent to $\sim\sim\varphi$.\\
$3. $ & $\, \varphi$  is semantically equivalent to some formula $\Box\psi$.\\
$4. $ & $\, \varphi$  is equivalent to some formula $\sim \psi$.
\end{tabular}
}
\epr

\bpr
{\rm For deterministic programs $\pi$, the interpretation of the construct $\overline{\pi}_{ij}$ is the
property of `being entangled according to (the linear map denoted by) $\pi'$. More precisely,
for deterministic $\pi$, we have
$$\M \overline{\pi}_{ij}\M=\overline{\M \pi \M}_{ij}$$
where we use the notation $\overline{F}_{ij}$ introduced (for any linear map $F$) after Proposition 6 of Section 2.2.
}
\epr

\section{Proof theory for $LQP$}

\subsection{Axioms for single systems}

First, we  admit {\it all the axioms and rules} of {\bf classical
$PDL$}, except for the ones concerning tests $\varphi?$ and Kleene star\footnote{We skip the axioms
for iteration $\pi^*$ only because we chose not to include this construct in our logic. However, if one
adds $\pi^*$ to our syntax, the usual $PDL$ axioms for iteration are still sound, so they can be added to the
proof system.}
$\pi^*$. In particular,
we have the following rules and axioms.
\smallskip\par\noindent
{\bf Substitution Rule. }
 $\, \, \mbox{ From } \vdash \Theta \mbox{ infer } \vdash \Theta [p / \varphi]$
\smallskip\par\noindent
And the `normality' conditions for the dynamic modalities $[\pi]$:
\smallskip\par\noindent
{\bf Kripke Axiom. } $\, \,  \vdash \, [\pi](p \to q) \rightarrow ([\pi] p \rightarrow [\pi] q)$
\par\noindent
{\bf Necessitation Rule. } $\, \, \mbox{ From }  \vdash p \mbox{ infer } \vdash [\pi] p$
\smallskip\par\noindent
Considering $\Box p$, we introduce the following axioms:
\par\noindent
{\bf Test Generalisation Rule. } If the variable $q$ does not occur in $\varphi$ or $\psi$, then,\\
$\, \, \mbox{ from }  \vdash \, \varphi \rightarrow [q?]\psi  \mbox{ infer } \vdash \, \varphi \to \Box \psi$
\par\noindent
{\bf Testability Axiom. }  $\, \, \vdash  \Box p \rightarrow [q?]p$

\smallskip\par\noindent
Testability can be stated in its dual form by means of $\langle q?
\rangle p \rightarrow \Diamond p$ or, equivalently, as $\langle q? \rangle p \rightarrow \langle p? \rangle
\top$.
This dual formulation of Testability allows us to give a
straightforward interpretation:
if the property associated to $p$ can be actualised by a measurement
(yielding an output state satisfying
 $p$), then we can directly test the property $p$ (by doing a measurement for $p$). The Test Generalisation Rule encodes the fact that $\Box$ is a universal quantifier over all possible measurements.

\smallskip\par\noindent
Other $LQP$-axioms are:

\smallskip\par\noindent
\begin{tabular}{lll}
{\bf Partial Functionality.}  \, \,  &$ \vdash $& $\neg [p?]q \rightarrow  [p?] \neg q $\\
{\bf Adequacy.} &$ \vdash $&$ p \wedge q \rightarrow \langle p? \rangle q $
\\
{\bf Repeatability.}  \, \, &$ \vdash $&$ T(p) \rightarrow [p?] p $
\\
{\bf Proper Superpositions.} \, \, &$ \vdash $&$ \langle \pi \rangle \Box \Box p \rightarrow [\pi'] p $
\\
{\bf Unitary Functionality.} \, \, &$ \vdash $ & $\neg [U]q \leftrightarrow  [U] \neg q $
\\
{\bf Unitary Bijectivity 1.} \,  \, &$ \vdash $&$ p \leftrightarrow [U;U^{\dagger}] p $
\\
{\bf Unitary Bijectivity 2.} \, \, &$ \vdash $&$ p \leftrightarrow [U^{\dagger};U] p $
\\
{\bf Adjointness.} \, \,  &$ \vdash $&$ p \rightarrow [\pi] \Box \langle \pi^{\dagger}\rangle \Diamond p $
\end{tabular}

\medskip

\bpr
{\rm Testability is closed under conjunctions, weakest preconditions; $\Box$-sentences,
orthocomplements and strongest postconditions are testable:
\begin{itemize}
\item \,\, $\vdash \, \, T(p)\wedge T(q) \, \rightarrow \, T(p\wedge q)$
\item \,\, $\vdash \, \, T(p) \, \rightarrow \, T([\pi] p)$
\item \,\, $\vdash \, \, T(\Box p)$
\item \,\, $\vdash \, \, T(\sim p)$
\item \,\, $\vdash \, \, T(\pi [p])$
\end{itemize}
}
\epr

\par\noindent A formula $\varphi$ is called {\it testable} if the theorem
$$\vdash \, \, T(\varphi)$$
is provable in our system. Observe that this notion is proof-theoretic. However,
the above proposition gives us a purely syntactical way to check testability:

\smallskip
\par\noindent\textbf{Corollary}. Any formula of the formula of the form $\Box\varphi$,
$\sim \varphi$ or $\top$, or which can be obtained from these formulas using only conjunctions $\phi\wedge \psi$
and weakest preconditions $[\pi] \varphi$, is testable.

\medskip
\bpr {\bf (Quantum logic, weak modularity or quantum modus ponens).} $\, \,$
{\rm  All the axioms and rules of traditional Quantum
Logic are satisfied by our {\it testable} formulas. In
particular, from our axioms one can prove `Quantum Modus
Ponens'\footnote{This explains why the weakest precondition
$[\varphi?]\psi$ has been taken as the basic implicational connective
in traditional Quantum Logic, under the name of `Sasaki hook' and
denoted by $\varphi\stackrel{S}{\to}\psi$.}
$\varphi \wedge [\varphi?]\psi \leq \psi$. In its turn, this rule
is equivalent  to the condition known in quantum logic as weak modularity, which is stated as follows: \\
$\varphi \wedge (\sim \varphi \sqcup (\varphi \wedge \psi)) \leq \psi$.
}\epr

\medskip

\bth {\bf (Soundness and Completeness).}
{\rm All the other axioms above are sound.
Moreover, {\it if we eliminate from the syntax of our logic all the special
constants} (both propositional constants $\top_I$, $1$ and $+$, and program
constants $\top_I$, $CNOT$, $X$, $H$, $Z$, and so on), {\it then
there exists a complete proof system for (single-system) Hilbert spaces},
which includes the above axioms.\footnote{In addition, the system includes
two more axioms of a rather technical nature, namely Piron's `Covering Law' \cite{JP4}
and `Mayet's Condition' \cite{Mayet}. See \cite{BaltagSmets} for details.}
}\Eth

The {proof} of this theorem is given in our paper \cite{BaltagSmets}, and is based
on an extension of (Mayet's version \cite{Mayet} of) Sol\`er's Theorem \cite{Soler}, which is itself an
extension of Piron's Representation Theorem
for Piron
lattices \cite{JP3,JP4,AA}.

\medskip
\bpr
{\rm The formula $\pi[\varphi]$
expresses the {\it strongest testable postcondition} ensured
by executing
program $\pi$ on any state satisfying (precondition) $\varphi$. In other
words, for every {\it testable} $\psi$, we have
$$\pi [\varphi] \leq \psi \, \, \mbox{ iff } \varphi \leq [\pi]\psi$$
}\epr

\medskip

\bpr ({\bf Adjointness Theorem}).
{\rm For all {\it testable} formulas $\varphi, \psi$, we have
$$\varphi \perp \pi [\psi] \, \, \mbox{ iff } \, \, \pi^{\dagger} [\varphi] \perp \psi$$
}
\epr

\subsection{Axioms for compound systems}
\par\noindent
\textbf{Separation Axioms.} Every state is $N$-separated; if a state is both $I$-separated and $J$-separated, then it is also $N\setminus I$-separated, $I \cup J$-separated and $I \cap J$-separated:
$$\vdash \top_N$$
and
$$\vdash \top_I \wedge \top_J \to \top_{N \setminus I} \wedge \top_{I\cup J} \wedge \top_{I\cap J}$$

\smallskip\par\noindent
\textbf{Axioms for the trivial $I$-local program}. {The program $\top_I$ is the weakest $I$-local program}; that is,
$$\vdash \,\, I(\pi) \, \, \rightarrow \,\, \langle \pi \rangle p \leq \langle \top_I \rangle p $$
and
$$\vdash \, I(\top_I)$$

As an immediate consequence, we obtain the following corollary.

\smallskip\par\noindent
{\bf Corollary 4.} The formula $\top_I$ is the weakest $I$-local property; that is,
$$\vdash I(\top_I)$$
and
$$\vdash I(p) \to p \leq \top_I$$

{\it Proof}. By the definition of $I$-locality $I(\pi)$ of a program, it is easy to see that the identity program $id$ is $I$-local for every $I$. Applying the first part of the above axiom (for $\top_I$), we obtain $\top_I = <id>\top_I \, \leq \, <\top_I>\top_I$, from which we deduce that $\top_I =\top_I \, \wedge \, <\top_I>\top_I$. Applying the definition of $\varphi_I$, we conclude that $\top_I = (\top_I)_I$, and thus (by definition of $I$-locality $I(\varphi)$ of a sentence) we derive $I(\top_I)$. The second part of the corollary follows trivially from the definition of $I(p)$.

\smallskip\par
\noindent Syntactically, we define an  {\it `$I$-local state'} to be any sentence $\varphi$ such that
$$\vdash \, I(\varphi) \wedge \varphi\not=\bot \wedge
\left( I(p)\wedge \bot\not=p\leq \varphi \, \rightarrow \, p=\varphi\right) $$
for some $p$ not occurring in $\varphi$.
In other words, these are propositions that can be proved to be atoms of the lattice of (consistent) $I$-local properties.



\smallskip\par\noindent
\textbf{Local States Axiom.} {Testable local properties are `local states'} (in the above sense, that is, atomic local properties): if $I\not=N$,
$$\vdash \, \, T(p) \wedge I(p) \wedge I(q) \wedge \bot\not=q \leq p  \,\, \rightarrow \, \,  q=p$$



\smallskip\par\noindent
\textbf{Basic-State Testability Axiom}. Our basic local states $c_i, \overline{\pi_{ij}}$ are testable and local (for the appropriate subsystem). More precisely, if $i,j\in N$, $c \in \{0,1, +, -\}$
and $\pi$ is a {\it deterministic} program, then
$$\vdash \, \, T(c_i) \wedge I(c_I) \wedge T(\overline{\pi_{ij}}) \wedge \{i,j\}(\overline{\pi_{ij}})$$

As an immediate consequence of the last two axioms, all constants of the form $\vec{c}_I$
(with $\vec{c}\in \{0,1, +, -\}^{|I|}$) are (testable) $I$-local states; similarly, if $\pi$ is deterministic then $\overline{\pi_{ij}}$
is a (testable) $\{i,j\}$-local state.


The following corollary is another immediate consequence.

\smallskip\par\noindent
{\bf Corollary 5.} $\,\,$ $\sim \top_I = \bot$

\smallskip

{\it Proof.} by the Adequacy axiom, we have $1_I \wedge \top_N \, \leq \, <1_I ?>\top_N$. But $\top_N$ is the `always true' sentence, so we have $\neg <0_I?>\top_N =[0_I?]\bot_N =\, \sim 0_I = 1_I = 1_i \wedge \top_N \leq \, <1_1?>\top_N$. From this, we get
$\top_N = (<0_I?>\top_N \, \vee \, {<1_I?>\top_N})$. By Adequacy again, we always have $\top_N \leq \, {<0_I?>0_I} \, \leq {<0_I?>\top_I}$ (since $0_I$ is $I$-local, so $0_I \leq \top_I$) and, similarly, $\top_N \leq \, <1_I?>\top_I$. Putting these three together, we deduce $\top_N \leq \, (<0_I?>\top_I \vee \, <1_I?>\top_I)$. But by the Testability axiom (in its dual form), we have $<0_I?>\top_I \leq \, \Diamond \top_I$ and, similarly, $<1_I?>\top_I \leq \Diamond \top_I$. Hence we have $\top_N \leq (\Diamond \top_I \vee \Diamond \top_I) = \Diamond \top_I = <\top_I?>\top_N$, and thus $\sim \top_I = [\top_I?]\bot_N = \neg<\top_I?>\top_N\leq \neg \top_N = \bot_N =\bot$.

\medskip

To capture the fact that the lattice of local properties is {\it atomistic}, we accept the following inference rule.

\smallskip\par\noindent
\textbf{Local Atomicity Rule}. $\,\,$ {Local properties are unions of testable local properties (that is, of local states)}:
if $I\not=N$ and the variable $p$ does not occur in $\varphi$, $\psi$ or $\theta$, then \\
from \, \, \, $\vdash \,\,  \psi \wedge T(p_I)\wedge p_I\leq \varphi \, \rightarrow \, p_I\leq \theta$ \\
infer \, \, \, $\vdash \, \, \psi \wedge I(\varphi) \, \rightarrow \, \varphi \leq \theta $

\medskip

As a consequence of the above axioms and rules, we obtain the following corollary.

\smallskip\par\noindent
\textbf{Corollary}. $\,$ {For $I\not=N$, every local state is testable}.
In other words, if $I\not=N$ and $p$ does not occur in $\varphi$, then from
$$\vdash \, I(\varphi) \wedge \varphi\not=\bot \wedge \left( I(p)\wedge \bot\not=p\leq \varphi \,
\rightarrow \, p=\varphi\right) $$
we can infer $$\vdash \, T(\varphi) .$$





The following axioms state that $+_i$ and $-_i$ are proper superpositions of $0_i$ and $1_i$.

\smallskip\par\noindent
{\bf Proper Superposition Axioms.} $\, \,  \vdash +_i \to \Diamond 0_i \wedge \Diamond 1_i$ and
$ \, \,  \vdash -_i \to \Diamond 0_i \wedge \Diamond 1_i$.

\medskip
The next axiom expresses the above-mentioned property of linear operators on
${\cal H}$ of being uniquely determined by their values on all the
states
$\mid x\rangle_1\otimes\cdots\mid x\rangle_n$, with
$\mid x\rangle_i\in\{\mid0\rangle_i, \mid 1 \rangle_i, \mid +\rangle_i\}$.

\bigskip
\par\noindent
\textbf{Determinacy Axiom of Deterministic Programs.} $\, \,$
For
deterministic programs $\pi , \pi'$,

$$\vdash \, \bigwedge_{\vec{c} \in \{0,1,+\}^n} \,
\left( \, \pi (\vec{c}_N)=\pi'(\vec{c}_N)\,  \rightarrow \,
\pi(p)= \pi' (p) \, \right) $$

The next axiom is the central one of our system, capturing the computational essence of entanglement,
as a semantic counterpart of Proposition 6 of Section 2.

\medskip\par\noindent
{\bf Entanglement Axiom}. If $\pi$ is deterministic and $i\not=j$, then
$$\vdash \,\, T(p_i) \, \, \rightarrow \, \, p_i? (\overline{\pi}_{ij}) \, =_j \, \pi_{ij}(p_i)$$

\medskip
Before presenting out next axioms, we note some consequence of the previous ones.
First, as for testability, we can define a proof-theoretic notion of locality.
A formula $\varphi$ is {\it $I$-local} if $\vdash \, I(\varphi)$ is a theorem; similarly,
a program $\pi$ is {\it $I$-local} if $\vdash \, I(\pi)$ is a theorem.

\bpr
{\rm Any formula of the form $\varphi_I$ is always $I$-local. Any formula of the form
$\overline{\pi_{ij}}$ is $\{i,j\}$-local.
If $\varphi$ and $\psi$ are $I$-local formulas and $\pi$ is an $I$-local program,
then $\varphi \vee \psi$, $\varphi\wedge \neg \psi$
and $\varphi\wedge [\pi] \psi$ are $I$-local. If $\varphi$ is $I$-local and $\psi$ is $J$-local,
then $\varphi\wedge\psi$ is $I\cup J$-local.
} \epr

\bpr
{\rm If $\varphi$ is a {\it testable} $I$-local formula, then $\varphi?$ is an $I$-local program.
$\top_I$ is $I$-local. If $\pi$ and $\pi'$ are $I$-local, then $\pi\cup\pi'$ and $\pi; \pi'$ are
$I$-local.
}
\epr

\bpr
{\rm Local programs act locally. In other words,
$$\vdash \, I(\pi) \wedge p=_I q \, \rightarrow \, p=_{N\setminus I} \pi(p) =_I \pi(q)$$
}
\epr

\bpr {\rm Systems composed of identical parts are identical:
$$\vdash \, p=_I q \wedge p=_J q \, \rightarrow \, p =_{I\cup J} q$$
}
\epr

\bpr {\rm $\vdash \,\,\,  p_I\perp q \, \leftrightarrow \, p_I\perp q_I$
}
\epr

\bpr ({\bf Dual Local Atomicity Rule}).
{\rm
If $I\not=N$, $\varphi$ and $\theta$ are $I$-separated, and $p$ does not occur in $\varphi, \psi$ or $\theta$, then, from
$$\vdash \,\,  \psi \wedge T(p_I)\wedge p_I \perp \varphi \, \rightarrow \, p_I\perp \theta$$
infer
$$\vdash \,\,  \psi \wedge T(\varphi_I)\wedge T(\theta_I) \, \rightarrow \, \varphi =_I \theta $$
}\epr

{\it Proof}. By using the fact that $p_I\perp q \leftrightarrow p_I\perp q_I$ and the $I$-locality of $p_I$,
we can rewrite the assumption
as
$$\vdash \,\,  \psi \wedge T(p_I)\wedge p_i \leq (\top_I \wedge\sim \varphi_I) \, \rightarrow \, p_I\leq (\top_I \wedge\sim \theta_I)$$
Now assume $\psi \wedge T(\varphi_I)\wedge T(\theta_I)$. Then
the formula $\top_I \wedge \sim\varphi_I$ = $\top_I\wedge \neg (\top_I \wedge [\varphi_I ?]\bot)$ is $I$-local
(since $\varphi_I$ is testable $I$-local, so $\varphi_I ?$ is an $I$-local program, we have $\top_I \wedge [\varphi_I ?]\bot$
is $I$-local) and, similarly, $\top_I \wedge \sim \theta_I$ is $I$-local. So we can apply the Local Atomicity
Rule to get $ (\top_I \wedge \sim \varphi_I) \leq (\top_I \wedge \sim \theta_I)$. Applying orthocomplementation,
we have $ \sim (\top_I \wedge \sim \theta_I) \leq \sim (\top_I \wedge \sim \varphi_I)$.
From this we get
$$\theta_I = \sim \sim \theta_I = \bot \sqcup \sim \sim \theta_I =
\sim \top_I \sqcup \sim \sim \theta_I= \sim (\top_I \wedge \sim \theta_I)$$
$$\leq \sim (\top_I \wedge \sim \varphi_I)=
\sim \top_I \sqcup \sim \sim \varphi_I = \bot \sqcup \varphi_I = \varphi_I$$
However, by the Local States Axiom,
this then implies that $\theta_I=\varphi_I$ (since both are testable $I$-local with $I\not=N$, and thus they are local states).
Since both $\theta_I$ and $\varphi_I$ are $I$-separated, it follows that $\theta =_I \varphi$.

\medskip
\par\noindent
\bth {\bf ({Compatibility of Programs Affecting Different Qubits}).}
\rm{
If $I \cap J = \emptyset$ and $\pi, \pi'$ are deterministic, then
$$\vdash I(\pi)\wedge J(\pi') \,\, \rightarrow \, \, \pi ; \pi' (p)
=\pi'; \pi (p)$$
}
\Eth

{\it Proof}. This is an immediate application of the Determinacy Axiom above. By that axiom,
it is enough to show the required identity for all $p$ of the form $p= \vec{c}_N$, with $\vec{c}\in \{0,1,+\}^n$.
Using the fact that $I\cup (N\setminus (I\cup J))\subseteq N\setminus J$ and $J\cup (N\setminus (I\cup J))\subseteq N\setminus I$
(since $I\cap J=\emptyset$) and Proposition 19 (saying that local programs `act locally'), we can easily
show that
$$(\pi; \pi')(\vec{c}_N) =_{N\setminus (I\cup J)} c_N =_{N\setminus (I\cup J)} (\pi' ; \pi) (\vec{c}_N)$$
$$(\pi; \pi')(\vec{c}_N) =_I \pi(\vec{c}_N) =_I (\pi' ; \pi) (\vec{c}_N)$$ and
$$(\pi; \pi')(\vec{c}_N) =_J \pi'(\vec{c}_N) =_J (\pi' ; \pi) (\vec{c}_N)$$
Using Proposition 20,
we put these together to conclude that
$$(\pi; \pi')(\vec{c}_N) =_{I\cup J \cup (N\setminus (I\cup J))} (\pi' ; \pi) (\vec{c}_N);$$ that is, that $$(\pi; \pi;) (\vec{c}_N) = (\pi' ; \pi)(\vec{c}_N)$$

\medskip
\bpr {\bf ({Dual Entanglement}).} {\rm
If $\pi$ is deterministic and $i\not=j$, then
$$\vdash \,\, T(q_j) \, \, \rightarrow \, \, q_j? (\overline{\pi_{ij}}) \, =_i \, \pi_{ij}^{\dagger}(q_j)$$
}\epr

{\it Proof}. Assume $T(q_j)$ and we need to show that $q_j? (\overline{\pi_{ij}}) \, =_i \, \pi_{ij}^{\dagger}(q_j)$.
It is easy to see that both sides are $i$-separated (that is, $\leq \top_i$), and also that both
$(q_j? (\overline{\pi_{ij}}))_i$ and $(\pi_{ij}^{\dagger}(q_j))_i$ are testable (since they are local states),
so we are in the conditions of the Dual Local Atomicity Rule (Proposition 22) above. By that Proposition, to prove the above identity,
it is enough to show that
$$\vdash \, \, T(p_i) \wedge p_i \perp \pi_{ij}^{\dagger}(q_j) \, \rightarrow \, p_i \perp q_j? (\overline{\pi_{ij}})$$
To show this, let $p_i$ be such that $T(p_i)$ and $p_i \perp \pi_{ij}^{\dagger}(q_j)$.
By the Adjointness Theorem, we have then $\pi_{ij}(p_i)\perp q_j$, and thus $q_j? (\pi_{ij} (p_i))=\bot$.
By the previous Proposition (on Compatibility of Programs on Different Qubits),
we have
$$p_i? (q_j? (\overline{\pi_{ij}}))= (p_i?; q_j?) (\overline{\pi_{ij}})$$ $$ = (q_j? ; p_i?)(\overline{\pi_{ij}}) $$ $$ =
q_j? (p_i? (\overline{\pi_{ij}}))$$ $$ = q_j? (\pi_{ij} (p_i))$$ $$ =\bot$$ (where we have used the Entanglement Axiom).
So we get $p_i \perp q_j? (\overline{\pi_{ij}})$. (Thus, using the Dual Local Atomicity Rule, the desired conclusion
follows).

\medskip
\bpr
{\rm ({\bf Entanglement Preparation Lemma}).
$$\vdash \, \, \pi_{ij}(p_i) \perp q_j \, \rightarrow \, \overline{\pi_{ij}} \perp (p_i \wedge q_j)$$
}
\epr

{\it Proof}. From the hypothesis, we get $q_j \perp (\pi_{ij} (p_i))_j$, and thus
$(p_i \wedge q_j) \perp (\pi_{ij} (p_i))_j$, from which it follows that
$(p_i \wedge q_j) \perp [p_i?](\pi_{ij} (p_i))_j$ (using the fact that $p_i? (p_i \wedge q_j)=p_i \wedge q_j$,
by Adequacy). On the other hand, we have $\overline{\pi_{ij}}\leq [p_i?](\pi_{ij} (p_i))_j$
(since $p_i? (\overline{\pi_{ij}}) \leq (p_i? (\overline{\pi_{ij}}) )_j =(\pi_{ij} (p_i))_j$, by the Entanglement
Axiom), so we get  $(p_i \wedge q_j) \perp \overline{\pi_{ij}}$.

\medskip
\bth ({\bf Teleportation Property}).
{\rm If $i, j, k$ are distinct indices then

$$\vdash \, \, (\overline{\sigma_{jk}} ? ; \overline{\pi_{ij}} ?) (p_i) \, =_k \,  (\pi_{ij};{\sigma_{jk}})(p_i)$$
}\Eth

{\it Proof}. By the same argument as above, it is enough to prove
$$ \vdash T(q_k)\wedge q_k \perp (\pi_{ij};\sigma_{jk})(p_i)\to q_k \perp (\overline{\sigma_{jk}}?;\overline{\pi_{ij}}?)(p_i)$$

\par\noindent
To show this, let $q_k$  be such that $T(q_k)$ and $T(q_k) \perp (\pi_{ij};\sigma_{jk})(p_i)$. Then $q_k \perp \sigma_{jk} (\pi_{ij}(p_i))$, and, by the Adjointness Theorem, we have $\sigma^{\dagger}_{jk}(q_k) \perp \pi_{ij}(p_i)$. By Dual Entanglement, it follows that $q_k?(\overline{\sigma_{jk}}) \perp \pi_{ij}(p_i)$. By the Entanglement Preparation Lemma, we have $\overline{\pi_{ij}} \perp (q_k?(\overline{\sigma_{jk}})\wedge p_i)$. Hence we get
\smallskip\par\noindent
\begin{tabular}{ll}
$q_{k}?((\overline{\sigma_{jk}}?;\overline{\pi_{ij}}?)(p_i))$ & $= \,\, q_k?(\overline{\pi_{ij}}?(\overline{\sigma_{jk}}?(p_i)))$ \\
& $= \,\, \overline{\pi_{ij}}?(q_k?(\overline{\sigma_{jk}}?(p_i)))$\\
& $ =_{ijk} \,\, \overline{\pi_{ij}}?(q_k?(\overline{\sigma_{jk}})\wedge p_i)$ \\
& $  = \,\,\, \perp$
\end{tabular}
\smallskip\par\noindent


\par\noindent
(where we have used Theorem 4 on the Compatibility of Programs on Different Qubits). So we get
$q_k \perp (\overline{\sigma_{jk}} ? ; \overline{\pi_{ij}} ?) (p_i)$, as desired.

\medskip
\par\noindent\textbf{Corollary 6}. If $i,j, k$ are distinct,
$$\vdash \, \, \overline{\pi_{ij}} ? (p_i\wedge \overline{\sigma_{jk}}) \, =_k \, (\pi_{ij}; \sigma_{jk}) (p_i)$$

{\it Proof}. By the Repeatability Axiom, we have $\overline{\sigma_{jk}}? (p_i) \leq \overline{\sigma_{jk}}$. Assuming
${\overline{\sigma_{jk}}? (p_i)}{\not=} \bot$, we get $\overline{\sigma_{jk}}? (p_i) =_{jk} \overline{\sigma_{jk}}$
(since $\overline{\sigma_{jk}}$ is testable and $\{j,k\}$-local, and hence it is a local state) and also that
$\overline{\sigma_{jk}}? (p_i) =_i p_i$ (since `local programs act locally', by Proposition 19). Thus, we get
$\overline{\sigma_{jk}}? (p_i) =_{ijk} p_i \wedge \overline{\sigma_{jk}}$. Applying the $\{i,j\}$ local program
$\overline{\pi_{ij}}$,
we get $$\overline{\pi_{ij}} ? (p_i\wedge \overline{\sigma_{jk}}) =_{ijk}
\overline{\pi_{ij}} ? (\overline{\sigma_{jk}}? (p_i))$$ $$\quad \quad \quad \quad \quad =
(\overline{\sigma_{jk}} ? ; \overline{\pi_{ij}} ?) (p_i) \, $$ $$\quad \quad \quad \quad \quad =_k \, (\pi_{ij}; \sigma_{jk}) (p_i)$$
from which we get the desired conclusion.

\medskip

By a refinement of the proof of Teleportation Property, we can prove the following proof-theoretic version of Lemma 3 in Section 2.2.

\medskip
\bpr ({\bf Entanglement Composition Lemma}).
{\rm
For distinct indices
$i,j,k,l$, programs $\pi, \pi', \pi''$ and local $\{1\}$-programs $\sigma_1, \rho_1$
we have
$$\vdash \overline{\pi}_{ij} \wedge \overline{\pi'}_{kl}\rightarrow
[\sigma_j; \rho_k; \overline{\pi''}_{jk}?]
\overline{(\pi; \sigma_1; \pi''; \rho_1^{\dagger};\pi')}_{il}$$
}
\epr

\smallskip

The {\it domain} $dom(\varphi)$ of a map $\pi$ is defined as
$dom(\pi) := <\pi>\top$.

\medskip
\bth ({\bf Agreement Property}).
\rm{ If two $I$-local maps $\pi, \pi'$ have the same domain and
they separate the input-state, then
their output states agree on all non-$I$ qubits: that is, if $I\cap
J=\emptyset$ then for all deterministic programs $\pi, \pi'$ we have\\
$\vdash \, \, T(p) \wedge I(\pi)\wedge I(\pi')\wedge dom(\pi)=dom(\pi')\wedge \pi(p)\leq \top_I\wedge
\pi'(p) \leq \top_I \, \rightarrow \, \pi (p) =_{N\setminus I} \pi' (p)$.
}\Eth

{\it Proof}. Put
$\psi := T(p)\wedge I(\pi) \wedge I(\pi') \wedge dom(\pi)=dom(\pi')\wedge \pi(p)\leq \top_I
\wedge \pi'(p) \leq \top_I$, and assume that $\psi$ is true.
By definition, $\pi(p)$ is testable (since $\pi$ is deterministic, so
$\pi(p)=\pi[p]=\sim [\pi^{\dagger}]\sim p$, and every sentence of the form $\sim \psi$ is testable), and the
same is true for $\pi'(p)$. So we can use the Dual Local Atomicity Rule to prove the above identity.
Let now $q_{N\setminus I}$ be such that $T(q_{N\setminus I})$ and $q_{N\setminus I} \perp \pi (p)$. Then $(\pi ; q_{N\setminus I}?) (p) =\bot$.
By the Compatibility of Programs on Different Qubits, we get
$(q_{N\setminus I}? ;\pi) (p) =\bot$, that is $p \leq [q_{N\setminus I}] [\pi] \bot= [q_{N\setminus I}]
\neg z\langle \pi \rangle \top=
[q_{N\setminus}] \neg dom(\pi)$. But $dom(\pi)=dom (\pi')$, so $p \leq [q_{N\setminus I}] \neg dom (\pi')
= [q_{N\setminus I}] [\pi'] \bot$, that is,
$(q_{N\setminus I}?; \pi') (p) =\bot$. Working now in reverse, we again apply the Compatibility of Programs on Different
Qubits, obtaining $(\pi'; q_{N\setminus I}?) (p) =\bot$, that is, $q_{N\setminus I} \perp \pi' (p)$.
 So we have proved that
$$\vdash \, \psi \wedge T(q_{N\setminus I}) \wedge q_{N\setminus I} \perp \pi (p)\rightarrow q_{N\setminus I} \perp \pi' (p)$$
By now applying the Dual Local Atomicity Rule, we get
$$\vdash \, \psi \rightarrow \pi (p) =_{N\setminus I} \pi' (p) ,$$
which is, the desired conclusion.

\noindent\textbf{Characteristic Formulas}.
\par\noindent
In order to formulate our
next axioms
(which deal with special logic gates), we now give some characteristic formulas for binary states, considering two qubits indexed by $i$ and $j$.
\smallskip\par\noindent
\begin{tabular}{|l|l|}
\hline
States & Characteristic Formulas\\
 \hline
& \\
$\overline{\mid 00 \rangle_{ij}} = \overline{\diraco_i \otimes \diraco_j}$ & $\langle 0_i ? \rangle 0_j \wedge [1_i ?] \perp$
 \\

 \hline
 Bell states: & \\
 $\beta_{xy}^{i,j} = \overline{\mid 0\rangle_i\otimes \mid y \rangle_j
+ (-1)^{x} \mid 1 \rangle_i\otimes \mid \tilde{y}\rangle_j}$ &  $\langle 0_i? \rangle y_j \wedge \langle 1_i ? \rangle \tilde{y}_j \wedge \langle +_i ? \rangle (-)^{x}_j$  \\
 with $\tilde{0} = 1$ and $\tilde{1} = 0$ , $x,y \in \{0,1\}$ & where $(-)^{x} = -$ if $x = 1$ \\
& and $(-)^{x} = +$ if $x = 0$
 \\
 \hline
 $\gamma^{i,j} = \beta_{00}^{i,j} + \beta_{01}^{i,j} =$ & \\
 $ \overline{\mid 00 \rangle_{ij} + \mid 01 \rangle_{ij} + \mid 10 \rangle_{ij} + \mid 11 \rangle_{ij} }$ & $\langle 0_i ? \rangle +_j \wedge \langle 1_i ? \rangle +_j \wedge \langle +_i ? \rangle +_j$\\
 \hline
 \end{tabular}

\bigskip\par\noindent
\textbf{Locality Axiom for Quantum Gates}.
\par\noindent
{Our special quantum gates are local, affecting only the specified qubits}:
$$\vdash \, \{i\} (X_i) \wedge \{i\} (Z_i) \wedge \{i\}(H_i) \wedge \{i,j\}(CNOT_{ij})$$
In addition to this, we require for $X, Z, H$.
\medskip\par\noindent
\textbf{Characteristic Axioms for Quantum Gates $X$ and $Z$.}
\par\noindent
\begin{eqnarray*}
\begin{array}{lllllllllll}
& \vdash & 0_i \to [X_i]1_i  & &
& \vdash & 1_i \to [X_i]0_i& &
& \vdash & +_i \to [X_i]+_i \\
& \vdash & 0_i \to [Z_i]0_i  & &
& \vdash & 1_i \to [Z_i]1_i  & &
& \vdash & +_i \to [Z_i]-_i \\
& \vdash & 0_i \to [H_i]+_i  & &
&  \vdash & 1_i \to [H_i]-_i & &
&  \vdash &+_i \to [H_i]0_i
\end{array}
\end{eqnarray*}
\noindent\textbf{Notation (Bell formulas)}
\par\noindent
For $x, y\in \{0,1\}$ and distinct indices
$i,j\in N$, we make the abbreviations for
$\beta_{xy}^{ij} := \overline{(Z_1^x ; X_1^y)}_{ij}$, and refer to these expressions as `the {\it Bell formulas}'.

\bpr
{\rm The Bell states  $\beta_{xy}^{i,j}$ are characterised by the
logic Bell
formulas $\beta_{xy}^{ij}$. In other words, a state satisfies one of
these formulas iff it coincides with the corresponding Bell state.
}
\epr

\smallskip

{\it Proof}. It is enough to check that the formulas $\beta_{xy}^{ij}$
imply the corresponding characteristic formulas in the above
table. For this, we use the Entanglement Axiom and the following (easily checked) theorems:
$$\vdash \, 0_1 \leftrightarrow  < Z_1^x; X_1^y > y_1$$
$$\vdash \, 1_1 \leftrightarrow  < Z_1^x; X_1^y > \tilde{y}_1$$
$$\vdash \, +_1 \rightarrow  < Z_1^x; X_1^y > (-)^x_1$$

\bigskip\par\noindent
{\bf Generalised Bell formulas, GHZ States.} As shown by the first author's student Dmitri
Akatov in his Master's thesis \cite{Akatov}, the above dynamic-logical characterisation
of Bell states can be recursively extended to the so-called {\it generalised (k-qubit) Bell states} (which form an orthonormal basis for the k-qubit space), for all $k \leq n$. Here, we only
mention a special case, that of the so-called $GHZ$ state (after Greenberg, Horne and
Zeilinger):
$$\beta_{000}^{i,j,k} = \, \overline{\mid 000>_{ijk} + \mid 111>_{ijk}}$$
\par
This state, of a special significance for various quantum protocols, can be characterised by the formula
$$\beta_{000}^{ijk} := <0_i ?>(0_j \wedge 0_k) \wedge <1_i ?>(1_j \wedge 1_k)\wedge <+_i ?>\beta_{00}^{jk}$$
From this, it is obvious that we have $+_i?(\beta_{000}^{ijk}) =_{jk} \beta_{00}^{jk}$; but one can easily check that we
also have $-_i ?(\beta_{000}^{ijk}) =_{jk} \beta_{10}^{jk}$. Using the notation $(-)^{z}$ introduced above for $z=0,1$ (putting $(-)^z := -$ if $z=1$ and $(-)^z := +$ if $z=0$), we can summarize this as
$$(-)^{z}_{i} ?(\beta_{000}^{ijk}) =_{jk} \beta_{z0}^{jk}$$

\smallskip
\par\noindent
\textbf{Characteristic Axioms for $CNOT$}.
$\, \,$ With the above notation, we put
\begin{eqnarray*}
\begin{array}{lllllll}
& \vdash & 0_i \wedge c_j \to  [CNOT_{ij}] c_j & &
& \vdash & 1_i \wedge 0_j \to   [CNOT_{ij}]1_j \\
& \vdash & 1_i \wedge 1_j \to   [CNOT_{ij}] 0_j & &
& \vdash &  1_i \wedge +_j \to  [CNOT_{ij}] +_j \\
& \vdash & +_i \wedge 0_j \to [CNOT_{ij}] \beta_{00}^{ij} & &
& \vdash & +_i \wedge 1_j \to   [CNOT_{ij}]\beta_{01}^{ij} \\
& \vdash & +_i \wedge +_j \to   [CNOT_{ij}]\gamma^{ij}
&{\mbox where } & &\gamma^{ij} &= \langle 0_i ? \rangle +_j \wedge \langle 1_i ? \rangle +_j \wedge \langle +_i ? \rangle +_j
\end{array}
\end{eqnarray*}
\bpr
For all $x,y \in \{0,1\}$,
$$\vdash \, (H_i ; CNOT_{i,j} (x_i \wedge y_j) = \beta_{xy}^{ij}$$
\epr
\noindent\textbf{Corollary 7}. If $i, j, k$ are all distinct,
$$ \, \, \vdash (CNOT_{ij}; H_j ; (x_i \wedge y_j) ? ) (p)=_k
\beta_{xy}^{i,j} ? (p)$$

{\it Proof}. From the above Proposition and from $H^{\dagger} = H$, $CNOT^{\dagger} = CNOT$, we get
$(CNOT_{i,j}; H_i) (\beta_{xy}^{ij}) = x_i \wedge y_i$, and thus

$$dom(CNOT_{i,j} ; H_i) = \langle CNOT_{ij}; H_i ; (x_i \wedge y_j) ? \rangle \top $$
$$ = \langle \beta_{xy}^{ij} ? \rangle \top$$
$$ = dom (\beta_{xy}^{ij}?)$$
The conclusion follows from this, together with the Agreement Property.

\bth {\rm All the above axioms and rules are sound for (quantum frames associated to) $n$-dimensional Hilbert spaces of the form $H^{\otimes n}$, where $H$ is any two-dimensional Hilbert space.
}
\Eth

The problem of obtaining a {\it complete} proof system for this logic is still open\footnote{However, we have strong reasons to believe the above system is {\it not} complete. At least one other sound
interesting axiom (of particular significance to quantum computing) has been proposed by the first author's student D. Akatov in his Master's thesis \cite{Akatov}. This is the `Determinacy of States' axiom, which
captures the converse of our Entanglement axiom: any entangled state is `entangled according to some
quantum program' $\pi$ (that is, it is of the form $\overline{\pi}_{ij}$); we chose not to include it here, as we have not used it in
this paper.}.
\section{Applications: correctness of quantum programs}
As applications to our logic, one can provide {\it formal correctness proofs} for a whole
range of quantum programs; one could claim that {\it all} quantum circuits and protocols
{\it in which probabilities do not play an essential role} can, in principle, be verified using
our logic, or some trivial extension of this logic (obtained by introducing more basic
constants for other relevant states and programs). In particular, all the quantum programs
covered by the `entanglement networks' approach in \cite{Coecke2} can be treated in this
logic. In his Master's thesis \cite{Akatov}
, D. Akatov has applied our logic to the
verification of various other protocols, for example, {\it superdense coding, quantum secret
sharing, entanglement swapping, logic gate teleportation, circuits for parallel computation of (sequential) compositions of programs using Bell base measurements}. The proofs are
modular, using as ingredients the main lemmas proved above: the Compatibility Theorem,
the Teleportation Property, the Entanglement Composition Lemma and the Agreement
Property. For simplicity, we will only consider two basic examples here: quantum
teleportation and quantum secret sharing.

\bigskip
\par\noindent
{\it Quantum teleportation}
\smallskip\par\noindent
Following \cite{Nielsen}, quantum teleportation is the name of a technique that makes it possible to `teleport'(that is, move) a quantum state between two agents, even in the absence of a quantum communication channel\footnote{However, note that a {\it classical} communication channel {\it is} required!} linking the sender and the recipient. We are working in $H\otimes H\otimes H$, with $H$ being the
two-dimensional (qubit) space, and so $n=3$.
There are two agents, Alice and Bob who, separated in space,
each having one qubit of an entangled EPR pair, represented by
$\beta_{00}^{2,3} \in H^{(2)}\otimes H^{(3)}$.  In addition to her part of the EPR pair, Alice  has another qubit $1$, in an unknown state
$q_1$. (Note that $q_1$ is a testable $1$-local property, since it is a $1$-local state.)
Alice wants to `teleport' this unknown qubit to Bob, that is, to execute a
program that will output a state satisfying $id_{13} (q_1)$. To do this, she
entangles the qubit $q_1$ with her part $q_2$ of the EPR pair, by performing first a
$CNOT_{1,2}$ gate on the two qubits and then a Hadamard transformation
$H_1$ on the first component.  Then Alice measures her qubits in the standard basis, thus destroying
the entanglement, so that Bob's qubit is left in a separated state $q_3$.
Though this state is
unknown, the results of Alice's measurements indicate the actions that Bob will have to
perform in order to transfer his qubit from state q3 into the state $id_{13} (q_1)$ (corresponding to the qubit
Alice had before the protocol).  It is thus
enough for Alice to send Bob two classical bits encoding the
result $x_1$ of the first measurement and the result $y_2$ of the
second measurement. To achieve `teleportation', Bob will have to apply the $X$-gate $y$ times, then apply the $Z$ gate $x$ times.

In our syntax, the quantum program described here is
$$\pi = \bigcup_{x,y\in \{0,1\}} CNOT_{1 2} ; H_1; (x_1\wedge y_2)?;
X_3^y ; Z_3^x $$
and the validity expressing the correctness of teleportation is
$$\vdash \, \pi (q_1\wedge \beta_{00}^{2,3}) =_3 id_{13} (q_1)$$
\par\noindent
To show this, observe that by applying the above Corollary to Proposition 27, in
which we take $i=1, j=2, k=3$,
we get that the validity above (to be proved) is equivalent to
$$\, \, \vdash\, (\beta_{xy}^{12}?; X_3^y; Z_3^x) (q_1\wedge \beta_{00}^{2,3}) =_3 id_{13} (q_1)$$
Replacing the logical Bell formulas with their definitions
$\beta_{xy}^{ij} := \overline{(Z_1^x ; X_1^y)}_{ij}$, we obtain the
following equivalent validity:
$$ \vdash \, (\overline{ (Z_1^x; X_1^y)_{12}}?; X_3^y; Z_3^x)(q_1\wedge \overline{id}_{23}) =_3
 id_{13} (q_1) , $$
where $id=Z_1^0; X_1^0$ is the identity. This last validity follows
from applying the (Corollary of) Teleportation Property and the validity $Z_1^x; X_1^y ; X_1^y; Z_1^x =id$ (due to
$X^{-1}=X, Z^{-1}=Z$).

\bigskip
\smallskip
\par\noindent
{\it Quantum secret sharing}
\medskip\par\noindent
As described in \cite{Gruska}
, this protocol realises the {\it splitting of quantum information}
into a given number $m$ of `shares' (among $m$ agents), such that the original information
(the `secret') can be recovered only by pooling together the information in all the shares.
The protocol uses $GHZ$ states in a similar way to teleportation. We consider here the
case $m = 3$ as an example: suppose Alice, Bob and Charles share a $GHZ$ triple state $\beta_{000}^{2,3,4}$ (each `having' one of the three entangled qubits, in increasing order: for example, Alice
has qubit 2, and so on). In addition, Alice has another qubit $1$, in an unknown state $q$.
To split this information $q_1$ into three shares, Alice measures her two qubits $1$ and $2$ in
the Bell basis, obtaining two bits $x, y$ (corresponding to which of the four Bell states $\beta_{xy}^{12}$ she obtained). After that, Bob measures his qubit $2$ in the dual basis $\{+, -\}$, obtaining
another bit $|->^{z}$ (with $z \in \{0, 1\}$)\footnote{Here we use for vectors a similar notation to the notation $(-)^{z}$ introduced for states in the previous section, that is, $\mid->^z := |->$ for $z =0$, and $|->^z := |+>$ for $z=1$.}. Finally, Charles is given qubit $4$, which is now in one
of 8 possible states $\psi_{4}^{(x,y,z)}$ (depending on the results obtained by Alice and Bob).

To recover the original `secret' $q$ from his qubit $\psi_{4}^{(x,y,z)}$, Charles can now apply a local
unitary transformation $Z_{4}^{z};X_{4}^{y};Z_{4}^{x}$. But notice that for this, he needs to know $x, y$ and $z$,
that is, the three agents have to share their information in order to recover $q$.

The quantum program described here is

$$\pi = \bigcup_{x,y \in \{0,1\}}\beta_{xy}^{12}?;(-)_{3}^{z}?;Z_{4}^{z};X_{4}^{y};Z_{4}^{x}$$

To prove correctness, we need to show
$$\vdash \pi(q_1 \wedge \beta_{000}^{234}) =_4 id_{14}(q_1)$$
\par\noindent
To show this, we use Compatibility and the $3$-locality of $(-)_{3}^{z}$
 to compute

$$(\beta_{xy}^{12}?;(-)_{3}^{z}?;Z_{4}^{z};X_{4}^{y};Z_{4}^{x})(q_1 \wedge \beta_{000}^{234}) = ((-)_{3}^{z};\beta_{xy}^{12}?;Z_{4}^{z};X_{4}^{y};Z_{4}^{x})(q_1 \wedge \beta_{000}^{234})$$
$$\quad \quad \quad \quad =(Z_{4}^{z};X_{4}^{y};Z_{4}^{x})(\beta_{xy}^{12}?(q_1 \wedge (-)_{3}^{z}?(\beta_{000}^{234})))$$

But recall that
$$(-)_{3}^{z}?(\beta_{000}^{234}) =_{24} \beta_{z0}^{24} =_{24} \overline{Z_{1}^{z}}_{24},$$
\par\noindent
so we have
$$\pi(q_1 \wedge \beta_{000}^{234}) =_{24}(Z_{4}^{z};X_{4}^{y};Z_{4}^{x})(\beta_{xy}^{12}?(q_1 \wedge \overline{(Z_{1}^{z})}_{24}))$$
$$\quad \quad \quad \quad \quad =(Z_{4}^{z};X_{4}^{y};Z_{4}^{x})(\overline{(Z_{1}^{x};X_{1}^{y})}_{12}?(q_1\wedge \overline{(Z_{1}^{z})}_{24}))$$
Applying the (Corollary of) the Teleportation Property, we get
\smallskip\par\noindent
\begin{tabular}{ll}
$\pi(q_1 \wedge \beta_{000}^{234}) $ & $=_4 (Z_{4}^{1};X_{4}^{y};Z_{4}^{x})((((Z_{1}^{x};X_{1}^{y}))_{12};(Z_{1}^{z})_{24})(q_1))$ \\
& $=(Z_{1}^{z};X_{1}^{y};Z_{1}^{x})_{14}((Z_{1}^{x};X_{1}^{y};Z_{1}^{z})_{14}(q_1)$ \\
& $=(Z_{1}^{x};X_{1}^{y};Z_{1}^{z};Z_{1}^{z};X_{1}^{y};Z_{1}^{x})_{14}(q_1)$ \\
& $= id_{14}(q_1)$
\end{tabular}
\medskip\par\noindent
\textbf{Note}: This proof can be easily generalised to the case of an $m$-share split (among $m$
agents) of the secret. See \cite{Akatov} for details.

\section{Conclusions and future work}

We have presented here a dynamic logic for compound quantum systems, capable
of {\it expressing and proving highly non-trivial features of quantum information flow,
such as entanglement and teleportation, properties of local transformations, separation, Bell states} and so on. The logic is Boolean, but has {\it modalities capturing the non-classical logical dynamics of quantum systems};
in addition,
it has {\it spatial features}, allowing us to express properties of {\it subsystems} of a compound quantum system.
The logic comes with a simple relational semantics, in terms of quantum states and quantum actions in a Hilbert space.
We have presented a sound proof system, which can be used to prove many interesting properties of quantum information,
including formal correctness proofs for a whole range of quantum protocols (we have
treated teleportation and quantum secret sharing here,
but there are also many others to be considered, such as superdense coding, entanglement
swapping and logic-gate teleportation).

However, a number of open problems remain. While in \cite{BaltagSmets} we sketched a completeness
result for the quantum dynamic logic of {\it single-system} quantum frames, {\it no corresponding completeness
result is known for compound systems}. So the completeness problem for the logic $LQP$ presented in this paper is still
open.

In this paper we have not included {\it iteration} (Kleene star) $\pi^*$ among our operations on programs, since it was not needed in our simple quantum programming applications. But one can, of course, add iteration and consider
the resulting logic, which would be useful in applications to quantum programs involving $while$-loops.
The usual $PDL$ axioms for Kleene star are sound, but, again, completeness remains an open
problem.

Another problem, which is of great importance for quantum computation, is
extending our setting to deal with the {\it quantitative aspects of quantum information} (in particular
with notions like phase and probability).
Our aim in this paper was to develop a logic to reason about
 {\it qualitative} quantum information flow, so we have ignored the {\it probabilistic} aspects
of quantum systems. There are natural ways to extend our setting, using quantum versions
of {\it probabilistic modal logic}, and we plan to investigate them in future work.


\medskip
\par\noindent\textbf{Acknowledgments.}
\par\noindent
We thank Bob Coecke for useful discussions, ideas and comments on this work,
and especially for presenting (a preliminary version of)
this paper at the 2nd International Workshop on Quantum Programming Languages (QPL2004) organised by Peter Selinger.
We thank Samson Abramsky and Prakash Panangaden for the useful discussions that took place at COMLAB in 2004 on the
topic of this paper. Sonja Smets thanks Amilcar Sernadas and Paulo Mateus for the useful discussions that took place at
IST in November 2004 on exogenous and dynamic quantum logic.

\small{

}

\end{document}